\documentclass[10pt,journal,compsoc]{IEEEtran}
\ifCLASSOPTIONcompsoc
  \usepackage[nocompress]{cite}
\else
  \usepackage{cite}
\fi
\ifCLASSINFOpdf
\else
\fi
\usepackage{amsmath}
\usepackage{url}
\usepackage{tikz}
\usetikzlibrary{shapes,arrows}
\usepackage{pgfplots}
\pgfplotsset{compat=1.13}
\usepackage{subcaption}

\usepackage{alphalph}

\usepackage{cases}
\newcommand{\teaser}[1]{\renewcommand{\vgtc@teaser}{\setlength{\captionmargin}{0.33in} #1 \setlength{\captionmargin}{0in}}}

\begin{document}
	
\title{A Visually Plausible Grasping System for Object Manipulation and Interaction in Virtual Reality Environments}

\author{Sergiu~Oprea,
        Pablo~Martinez-Gonzalez,
        Alberto~Garcia-Garcia,
        John~Alejandro~Castro-Vargas,
        Sergio~Orts-Escolano, 
        and~Jose~Garcia-Rodriguez% <-this % stops a space
\IEEEcompsocitemizethanks{\IEEEcompsocthanksitem Sergiu Oprea, Pablo M. Gonzalez, Alberto G. Garcia, John Alejandro C. Vargas, Sergio O. Escolano, and Jose G. Garcia are with the 3D Perception Lab at the University of Alicante, Spain.\protect\\
E-mails: soprea@dtic.ua.es, pmartinez@dtic.ua.es, agarcia@dtic.ua.es, jacastro@dtic.ua.es, sorts@ua.es, jgarcia@dtic.ua.es.}% <-this % stops an unwanted space
\thanks{Manuscript received April 19, 2005; revised August 26, 2015.}}


\markboth{Journal of \LaTeX\ Class Files,~Vol.~14, No.~8, August~2015}%
{Shell \MakeLowercase{\textit{et al.}}: Bare Demo of IEEEtran.cls for Computer Society Journals}




\IEEEtitleabstractindextext{%

\begin{abstract}

Interaction in virtual reality (VR) environments is essential to achieve a pleasant and immersive experience. Most of the currently existing VR applications, lack of robust object grasping and manipulation, which are the cornerstone of interactive systems. Therefore, we propose a realistic, flexible and robust grasping system that enables rich and real-time interactions in virtual environments. It is visually realistic because it is completely user-controlled, flexible because it can be used for different hand configurations, and robust because it allows the manipulation of objects regardless their geometry, i.e. hand is automatically fitted to the object shape. In order to validate our proposal, an exhaustive qualitative and quantitative performance analysis has been carried out. On the one hand, qualitative evaluation was used in the assessment of the abstract aspects such as: hand movement realism, interaction realism and motor control. On the other hand, for the quantitative evaluation a novel error metric has been proposed to visually analyze the performed grips. This metric is based on the computation of the distance from the finger phalanges to the nearest contact point on the object surface. These contact points can be used with different application purposes, mainly in the field of robotics. As a conclusion, system evaluation reports a similar performance between users with previous experience in virtual reality applications and inexperienced users, referring to a steep learning curve.
\end{abstract}

}

\maketitle

\IEEEdisplaynontitleabstractindextext

% \fi
\IEEEpeerreviewmaketitle

\IEEEraisesectionheading{\section{Introduction}\label{sec:introduction}}

\begin{figure*}[t]
	\centering
	\begin{subfigure}[t]{0.35\linewidth}
		\includegraphics[width=\linewidth]{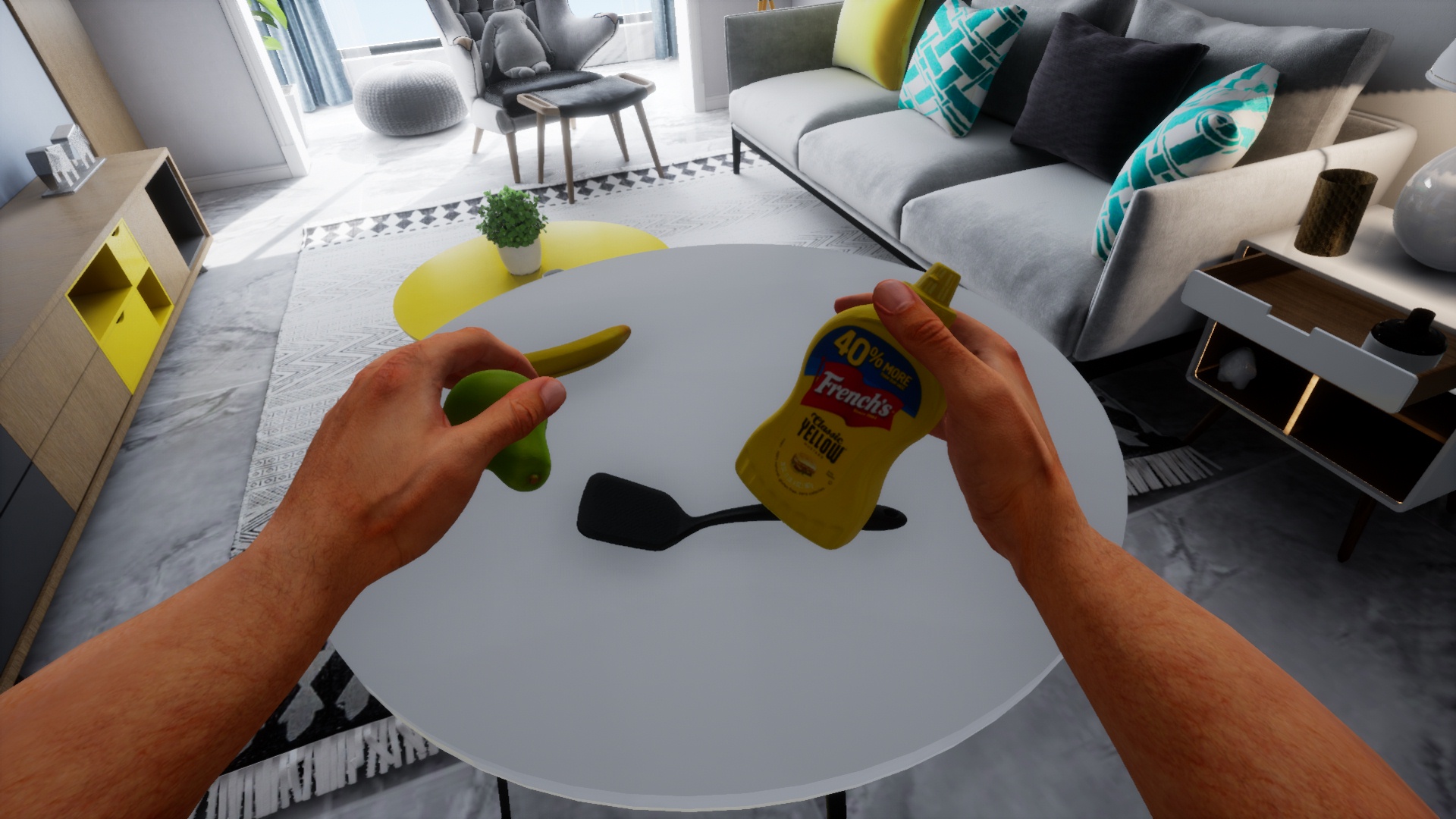}
		\label{fig:real0}
	\end{subfigure}
	\quad
	\begin{subfigure}[t]{0.35\linewidth}
		\includegraphics[width=\linewidth]{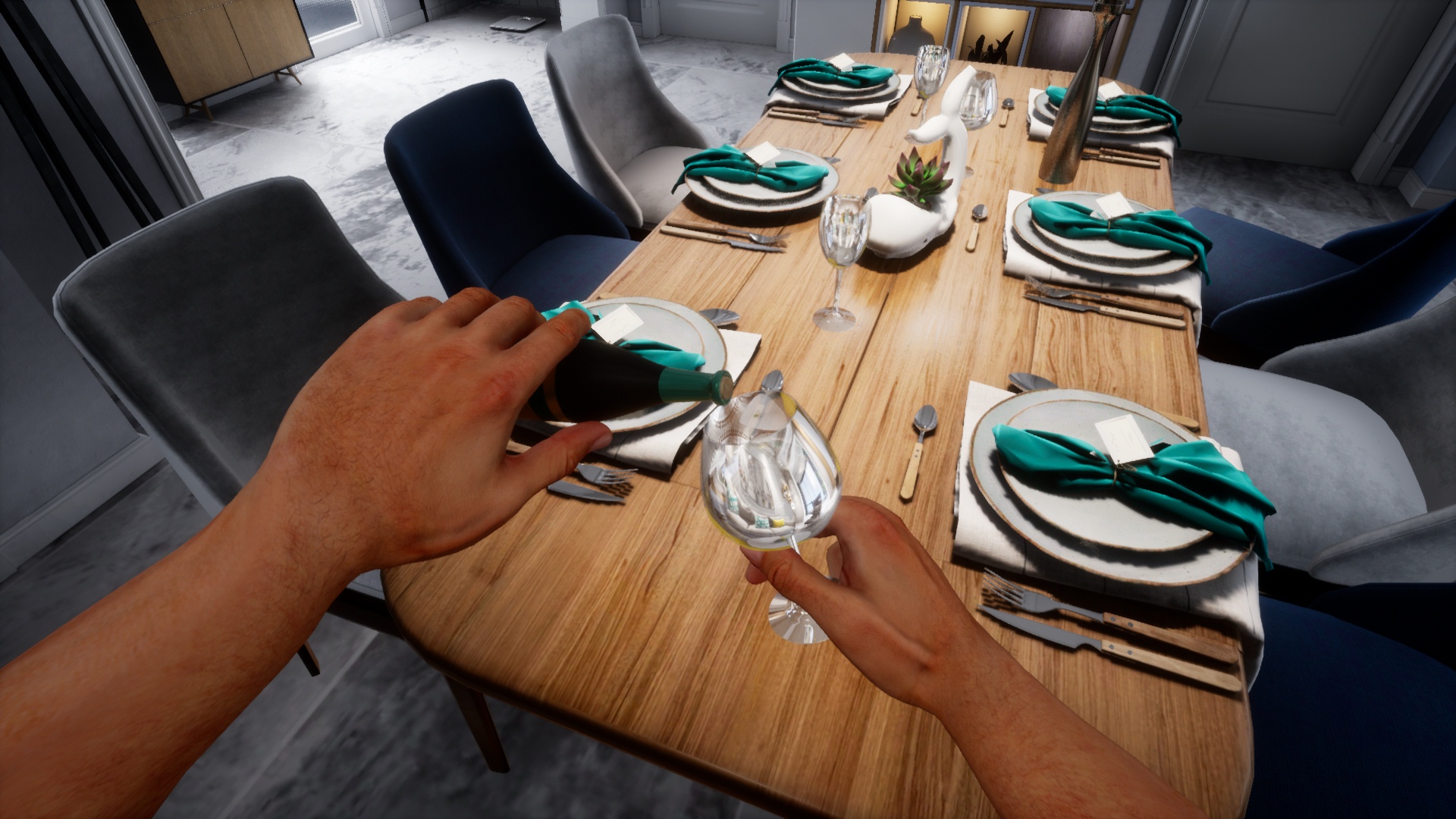}
		\label{fig:real1}
	\end{subfigure}
	\quad
	\begin{subfigure}[t]{0.35\linewidth}
		\includegraphics[width=\linewidth]{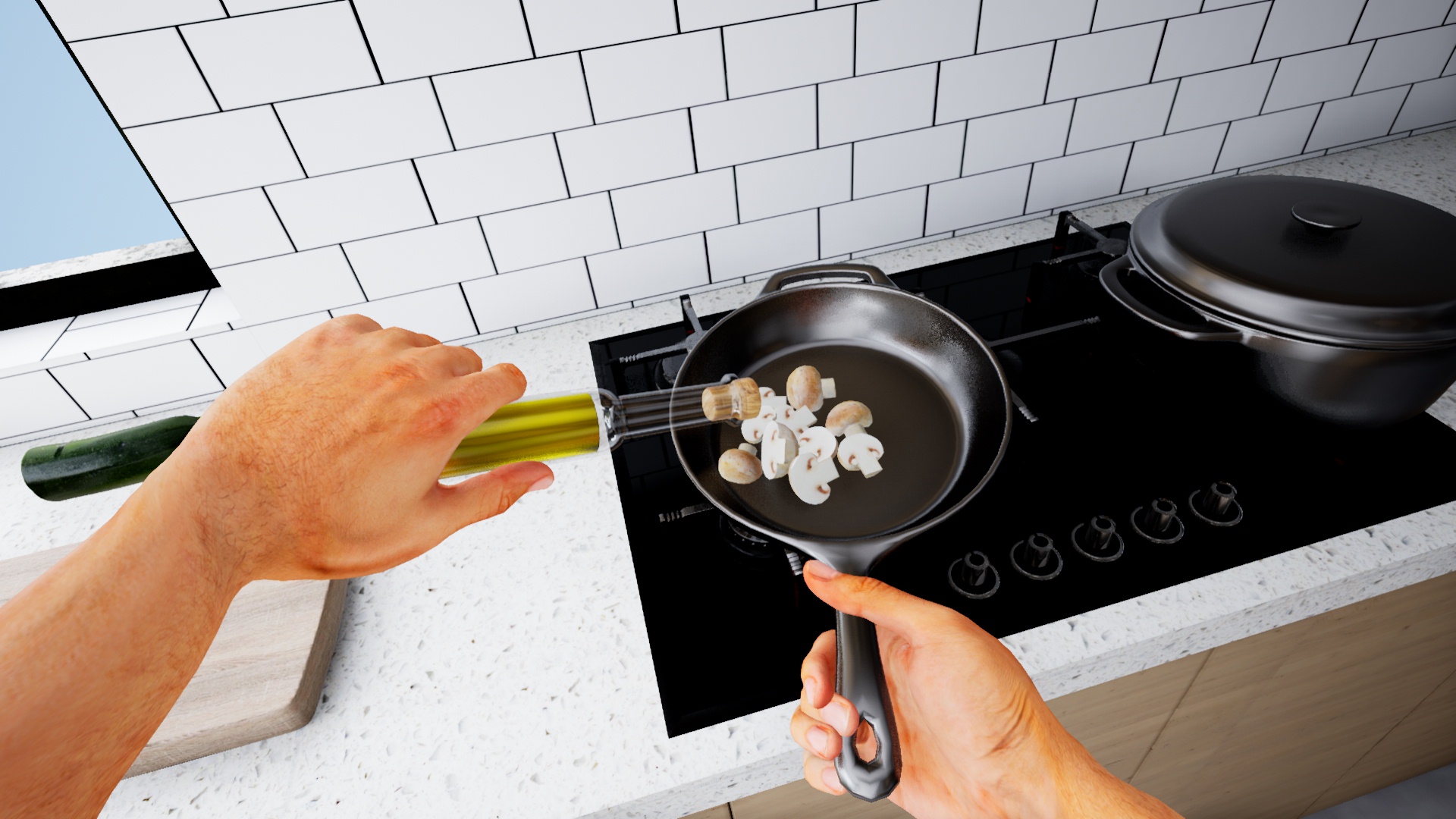}
		\label{fig:real2}
	\end{subfigure}
	\quad
	\begin{subfigure}[t]{0.35\linewidth}
		\includegraphics[width=\linewidth]{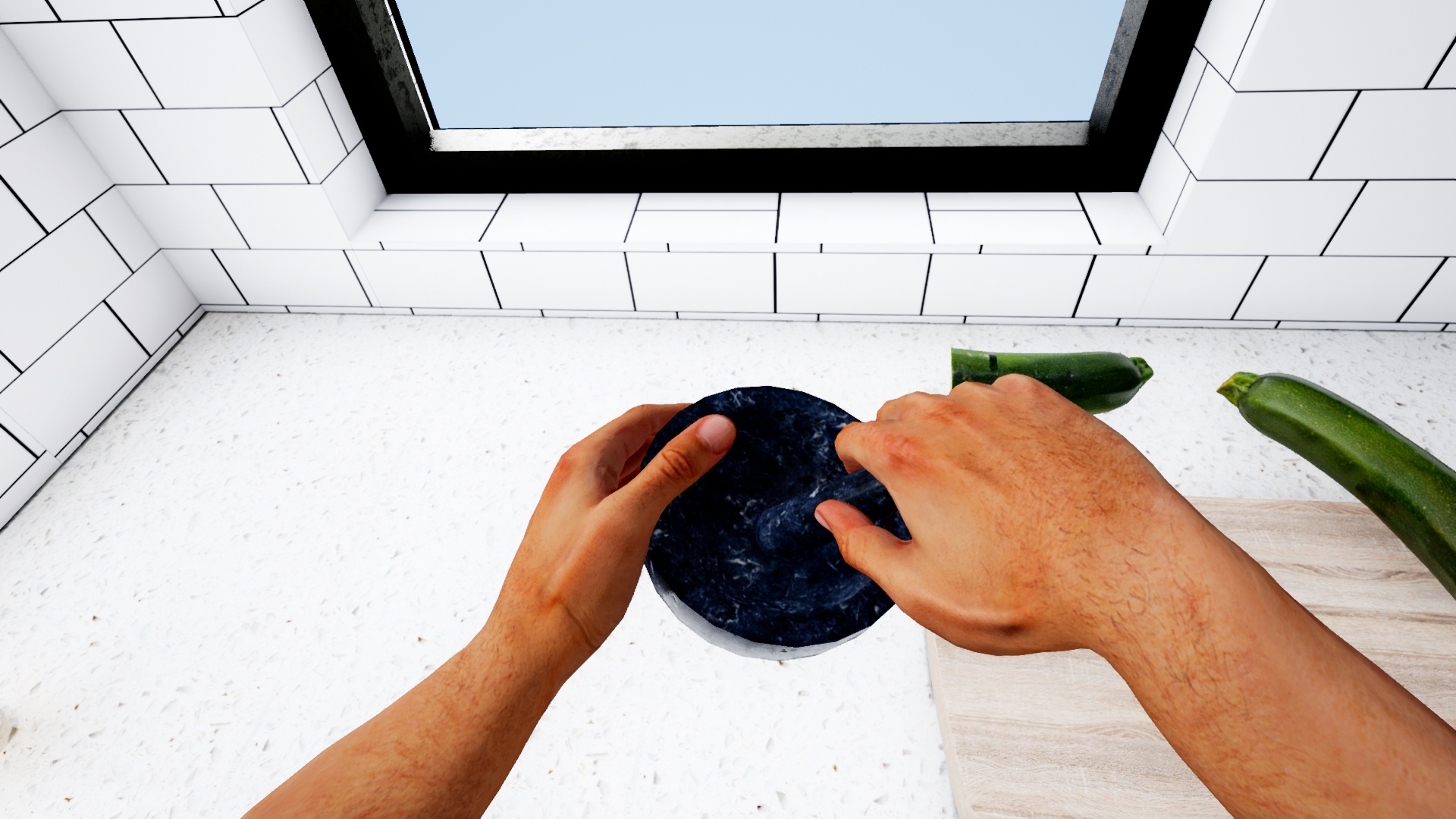}
		\label{fig:real3}
	\end{subfigure}
	\caption{Examples of interaction with objects extracted from the YCB dataset \cite{calli2015} in a photorealistic virtual environment, where the user: grabs a pear and mustard (top left), serves wine in a glass (top right), cooks mushrooms in a frying pan with a little oil (bottom left), and uses the mortar (bottom right).}
	\label{fig:interaction}
\end{figure*}


\IEEEPARstart{W}{ith} the advent of affordable VR headsets such as Oculus VR/Go and HTC Vive, many works and projects are using virtual environments for different purposes. Most of VR applications are related to the entertainment industry (i.e. games and 3D cinema) or architectural visualizations, where virtual scene realism is a cornerstone. Currently existing VR systems are limited by their resolution, field-of-view, frame rate, and interaction among other technical specifications. In order to enhance user VR experience, developers are also focused on implementing rich interactions with the virtual environment, allowing the user to explore, interact and manipulate scene objects as in the real world.

Interaction is a crucial feature for training/simulation applications (e.g. flight, driving and medical simulators), and also teleoperation (e.g. robotics), where the user ability to interact and explore the simulated environments is paramount for achieving an immersive experience. For this purpose, most of VR devices come with a pair of handheld controllers which are fully tracked in 3D space and specifically designed for interaction. One of the most basic interaction tasks is object grasping and manipulation. In order to achieve an enjoyable experience in VR, a realistic, flexible and real-time grasping system is needed. However, grasp synthesis in manipulation tasks is not straightforward because of the unlimited number of different hand configurations, the variety of object types and their geometries, and also due to the selection of the most suitable grasp for every different object in terms of realism, kinematics and physics.

Currently existing real-time approaches in VR are purely animation-driven, completely relying on the animations realism. Moreover, these approaches are constrained to a limited number of simple object geometries and unable to deal with unknown objects. For every different object type and geometry, predefined animations are needed. This fact hinders the user experience, limiting its interaction capabilities. For a complete immersion user should be able to interact and manipulate different virtual objects as in the real world.

In this paper, we propose a real-time grasping system for object interaction in virtual reality environments. We aim to achieve natural and visually plausible interactions in photorealistic environments rendered by Unreal Engine. Taking advantage of headset tracking and motion controllers, a human operator can be embodied in such environments as a virtual human or robot agent to freely navigate and interact with objects. Our grasping system is able to deal with different object geometries, without the need of a predefined grasp animation for each. With our approach, fingers are automatically fitted to object shape and geometry. We constrain hand finger phalanges motion checking in real-time for collisions with the object geometry.

Our grasping system was analyzed both qualitatively and quantitatively. On one side, for the qualitative analysis, grasping system was implemented in a photorealistic environment where the user is freely able to interact with real world objects extracted from the YCB dataset \cite{calli2015} (see Figure \ref{fig:interaction}). The qualitative evaluation is based on a questionnaire that will address the user interaction experience in terms of realism during object manipulation and interaction, system flexibility and usability, and general VR experience. On the other side, a quantitative grasping system analysis was carried out, contrasting the elapsed time a user needs in grasping an object and grasp quality based on a novel error metric which quantifies the overlapping between hand fingers and grasped object.

From the quantitative evaluation, we obtain individual errors for the last two phalanges of each finger, the time user needed to grasp the object and also the contact points. This information alongside other provided by UnrealROX \cite{martinez2018} such as depth mpas, instance segmentations, normal maps, 3D bounding boxes and 6D object pose (see Figure \ref{fig:unrealroxgt}), enables different robotic applications as described in Section \ref{sec:applications}.

In summary, we make the three following contributions:
\begin{itemize}
	\item We propose a real-time, realistic looking and flexible grasping system for natural interaction with arbitrary shaped objects in virtual reality environments; 
	\item We propose a novel metric and procedure to analyze visual grasp quality in VR interactions quantifying hand-object overlapping; 
	\item We provide the contact points extracted during the interaction in both local and global system coordinates.
\end{itemize}

The rest of the paper is structured as follows. First of all, Section \ref{sec:related_works} analyzes the latest works related to object interaction and manipulation in virtual environments. The core of this work is comprised in Section \ref{sec:graspingsystem} where our approach is described in detail. Then, the performance analysis, with the qualitative and our novel quantitative evaluations, is discussed in Section \ref{sec:experimentation}. Analysis results are reported in Section \ref{sec:results}. Then, several applications are discussed in Section \ref{sec:applications}. After that, limitations of our approach are covered in Section \ref{sec:limitationsfutureworks} alongside several feature works. Finally, some conclusions are drawn in the last Section \ref{sec:conclusion}.

\section{Related works}
\label{sec:related_works}

Computer graphics are fundamental for virtual reality applications, bringing realistic environments to users. However, currently existing virtual reality systems come with a pair of handheld devices specifically designed to enable user interaction with the virtual environment. This has led researchers to focus on designing efficient and realistic interactions with the virtual environment in order to improve the user experience. In spite of existing approaches, VR interaction remains an open problem for researchers and companies. 

Grasping action is the most basic component of any interaction and it is composed of three major components \cite{aydin1999}. The first one is related to the process of approaching the arm and hand to the target object, considering the overall body movement. The second component focuses on the hand and body pre-shaping before the grasping action. Finally, the last component fits the hand to the geometry of the object by closing each of the fingers until contact is established.

Our work and most of the currently existing works about virtual reality interaction and grasping, relate to the third component of the grasping action. Most of the currently existing approaches in solving this problem are data-driven. This is using predefined animations for concrete object geometries which are stored in a database. The keys of data-driven approaches are to effectively index large datasets in order to quickly match unknown object geometries with existing hand poses in the database.

In this section we will only mention the most recent works and approaches which are mostly aligned with our proposal and preferably related to virtual reality applications in which interaction is done by means of handheld devices, such as Oculus Touch controllers. Moreover and for an in-depth insight, a detailed review of the advances conducted in hand modeling and animation was published by Wheatland et al. \cite{wheatland2015}. At the same time and regarding 3D object selection techniques in virtual environments, a review was published by Argelaguet et al. \cite{argelaguet2013}.

\subsection{Data-driven approaches}

Grasping data-driven approaches have existed since a long time ago \cite{aydin1999}. These methods are based on large databases of predefined hand poses selected using user criteria or based on grasp taxonomies (i.e. final grasp poses when an object was successfully grasped) which provide us the ability to discriminate between different grasp types. 

From this database, grasp poses are selected according with given object shape and geometry \cite{li2007} \cite{goldfeder2011}. Li et al. \cite{li2007} construct a database with different hand poses and also object shapes and sizes. Despite having a good database, the process of hand poses selection is not straightforward since there can be multiple equally valid possibilities for the same gesture. To address this problem, Li et al. \cite{li2007} proposed a shape-matching algorithm which returns multiple potential grasp poses. 

The selection process is also constrained by the hand high degree of freedom (DOF). In order to deal with dimensionality and redundancy many researchers have used techniques such as principal component analysis (PCA) \cite{braido2004}\cite{ciocarlie2007}. For the same purpose, Jorg et al. \cite{jorg2009} studied the correlations between hand DOFs aiming to simplify hand models reducing DOF number. The results suggest to simplify hand models by reducing DOFs from 50 to 15 for both hands in conjunction without loosing relevant features.   

\subsection{Hybrid data-driven approaches}

In order to achieve realistic object interactions, physical simulations on the objects should also be considered \cite{pollard2005}\cite{kry2006}\cite{bai2014}. Moreover, hand and finger movement trajectories need to be both, kinematically and dynamically valid \cite{ye2012}. Pollard et al. \cite{pollard2005} simulate hand interaction, such as two hands grasping each other in the handshake gesture. Bai et al. \cite{bai2014} simulate grasping an object, drop it on a specific spot on the palm and let it roll on the hand palm. A limitation of this approach is that information about the object must be known in advance, which disable robot to interact with unknown objects. By using an initial grasp pose and a desired object trajectory, the algorithm proposed by Liu \cite{liu2009} can generate physically-based hand manipulation poses varying the contact points with the object, grasping forces and also joint configurations. This approach works well for complex manipulations such as twist-opening a bottle. Ye and Liu \cite{ye2012} reconstruct a realistic hand motion and grasping generating feasible contact point trajectories. Selection of valid motions is defined as a randomized depth-first tree traversal, where nodes are recursively expanded if they are kinematically and dynamically feasible. Otherwise, backtracking is performed in order to explore other possibilities. 

\subsection{Virtual reality approaches}

This section is limited to virtual reality interaction using VR motion controllers, avoiding glove-based and bare-hand approaches. Implementation of the aforementioned techniques in virtual reality environments is a difficult task cause optimizations are needed to keep processes running in real time. Most of current existing approaches for flexible and realistic grasping are not suitable for real-time interaction. VR developers aim to create fast solutions with realistic and natural interactions. 

Recent approaches are directly related to the entertainment industry, i.e. video games. An excellent example is \emph{Lone Echo}, a narrative adventure game which consists of manipulating tools and objects for solving puzzles. Hand animations are mostly procedurally generated, enabling grasping of complex geometries regardless their grasp angle. This approach \cite{copenhaver2017} is based on a graph traversal heuristic which searches intersections between hand fingers and object surface mesh triangles. A* heuristic find the intersection that is nearest to the palm and also avoid invalid intersections. After calculating angles to make contact with each intersection point, highest angle is selected and fingers are rotated accordingly. 

Mostly implemented solutions in VR are animation-based \cite{oculusDistanceGrab2017} \cite{oculusFirstExperience} \cite{vrtemplate2016}. These approaches are constrained to a limited number of simple object geometries and are unable to deal with unknown objects. Movements are predefined for concrete object geometries, hindering user interaction capabilities in the virtual environment. In \cite{oculusDistanceGrab2017}, distance grab selection technique is implemented to enhance the user comfort when interacting in small play areas, while sitting or for grabbing objects on the floor. Grasping system is based on three trigger volumes attached to each hand: two small cylinders for short-range grasp, and a cone for long-range grabbing. Based on this approach, we used trigger volumes attached to finger phalanges to control its movement and detect object collisions more precisely. In this way we achieve a more flexible and visually plausible grasping system enhancing immersion and realism during interactions.

\section{Grasping system}
\label{sec:graspingsystem}

With the latest advances in rendering techniques, visualization of virtual reality (VR) environments is increasingly more photorealistic. Besides graphics, which are the cornerstone of most VR solutions, interaction is also an essential part to enhance the user experience and immersion. VR scene content is portrayed in a physically tangible way, inviting users to explore the environment, and interact or manipulate represented objects as in the real world. VR devices aim to provide very congruent means of primary interaction, described as a pair of handheld devices with very accurate 6D one-to-one tracking. The main purpose is to create rich interactions producing memorable and satisfying VR experiences.

Most of the currently available VR solutions and games lack of a robust and natural object manipulation and interaction capabilities. This is because, bringing natural and intuitive interactions to VR is not straightforward, which makes VR development challenging at this stage. Interactions need to be in real-time and maintaining a high and solid frame rate, directly mapping user movement to VR input in order to avoid VR sickness (visual and vestibular mismatch). Maintaining the desired 90 frames per second (FPS) in a photorealistic scene alongside complex interactions is not straightforward. This indicates the need of a flexible grasping system designed to naturally and intuitively manipulate unknown objects of different geometries in real-time.

\subsection{Overview}
\label{subsec:overview}

Our grasping approach was designed for real-time interaction and manipulation in virtual reality environments by providing a simple, modular, flexible, robust, and visually realistic grasping system. Its main features are described as follows:
\begin{itemize}
	\item Simple and modular: it can be easily integrated with other hand configurations. Its design is modular and adaptable to different hand skeletals and models.
	\item Flexible: most of the currently available VR grasp solutions are purely animation-driven, thus limited to known geometries and unable to deal with previously unseen objects. In contrast, our grasping system is flexible as it allows interaction with unknown objects. In this way, the user can freely decide the object to interact with, without any restrictions.
	\item Robust: unknown objects can have different geometries. However, our approach is able to adapt the virtual hand to objects, regardless of their shape.
	\item Visually realistic: grasping action is fully controlled by the user, taking advantage of its previous experience and knowledge in grasping daily common realistic objects such as cans, cereal boxes, fruits, tools, etc. This makes resulting grasping visually realistic and natural just as a human would in real life.
\end{itemize}
The combination of the above described features makes VR interaction a pleasant user experience, where object manipulation is smooth and intuitive.

\begin{figure}
	\centering
	\includegraphics[width=.65\linewidth]{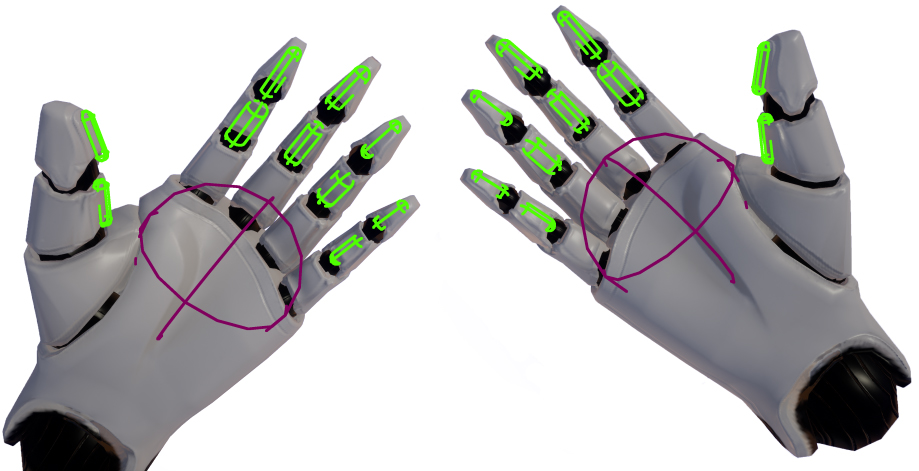}
	\caption{In green, capsule triggers of the middle and distal phalanges. In purple, sphere trigger of the palm.}
	\label{fig:mannequinhands}
\end{figure}

Our grasping works by detecting collisions with objects through the use of trigger actors placed experimentally on the finger phalanges. A trigger actor is a component from Unreal Engine 4 used for casting an event in response to an interaction, e.g. collision with another object. These components can be of different shapes, such as capsule, box, sphere, etc. In the Figure \ref{fig:mannequinhands} capsule triggers are represented in green and palm sphere trigger in red. We experimentally placed two capsule triggers on the last two phalanges of each finger. We noticed that this configuration is the most effective in detecting objects collisions. Notice that collision detection is performed for each frame, so, for heavy configurations with many triggers, performance would be harmed. 

\subsection{Components}

\definecolor{lightgreen}{RGB}{166,196,138}
\definecolor{lightblue}{RGB}{143,184,237}

\tikzstyle{block} = [rectangle, draw, fill=lightgreen, 
text width=5em, text centered, rounded corners, minimum height=3em]
\tikzstyle{cloud} = [draw, ellipse, text centered, fill=lightblue, text width=4em,
minimum height=2em]
\tikzstyle{line} = [draw, -latex']

\begin{figure}
	\centering
	\begin{tikzpicture}[node distance = 2.2cm, auto]
	% Place nodes
	\node [block] (init) {Object selection};
	\node [block, right of= init] (manager) {Interaction manager};
	\node [block, right of= manager] (movement) {Finger movement};
	\node [block, right of= movement] (logic) {Grasping logic};
	
	% Draw edges
	\path [line] (init) -- (manager);
	\path [line] (manager) -- (movement);
	\path [line] (movement) -- (logic);
	\end{tikzpicture}
	\caption{Pipeline with grasping system components}
	\label{fig:pipeline}
\end{figure}
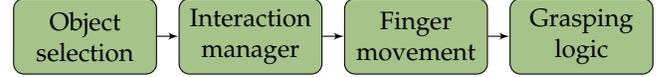

Our grasping system is composed of the components represented in the Figure \ref{fig:pipeline}. These components are defined as follows:
\begin{itemize}
	\item Object selection: selects the nearest object to the hand palm. Detection area is determined by the sphere trigger placed on the hand palm (red colored in Figure \ref{fig:mannequinhands}). The sphere trigger returns the world location of all the overlapped actors. As a result, the nearest actor can be determined by computing the distance from each overlapped actor to the center of the sphere trigger. Smallest distance will determine the nearest object, saving its reference for the other components.
	\item Interaction manager: manages capsule triggers which are attached to finger phalanges as represented in Figure \ref{fig:mannequinhands}. If a capsule trigger reports an overlap event, the movement of its corresponding phalanx is blocked until hand is reopened or the overlapping with the manipulated object is over. The phalanx state (blocked or in movement) will be used as input to the grasping logic component. A phalanx is blocked if there is an overlap of the its corresponding capsule trigger with the manipulated object.
	\item Finger movement: this component determines the movement of the fingers during the hand closing and opening animations. It ensures a smooth animation avoiding unexpected and unrealistic behavior in finger movement caused neither by a performance drop or other interaction issues. Basically, it monitors each variation in the rotation value of the phalanx. In the case of detecting an unexpected variation (i.e. big variation) during a frame change, missing intermediate values will be interpolated so as to keep finger movement smooth. 
	\item Grasping logic: this component manages when to grab or release an object. This decision is made based on the currently blocked phalanges determined with the interaction manager component. The object is grasped or released based on the following function:
\end{itemize}	
	%\setlength{\arraycolsep}{0.0em}
	%\begin{numcases} {\label{eq:3} f(th_{ph}, in_{ph}, mi_{ph}, palm)=}
	%true, & if $(th_{ph} \lor palm) \land (in_{ph} \lor mi_{ph})$\\
	%false, & otherwise
	%\end{numcases}, where function inputs are
	\begin{equation} \label{eq:3}
	f(x)=
	\begin{cases}
	true, & \text{if } (th_{ph} \lor palm) \land (in_{ph} \lor mi_{ph})\\
	false,              & \text{otherwise}
	\end{cases}
	\end{equation}, where $x = (th_{ph}, in_{ph}, mi_{ph}, palm)$ is defined as
	
	\begin{equation} \label{eq:4}
	\begin{array}{l}
	th_{ph} = thumb_{mid} \lor thumb_{dist} \\
	in_{ph} = index_{mid} \lor index_{dist} \\
	mi_{ph} = middle_{mid} \lor middle_{dist}
	\end{array}
	\end{equation}
	
	Equation \ref{eq:3} determines when an object is grasped or released based on the inputs determined in Equation \ref{eq:4} where $th_{ph}$, $in_{ph}$, and $mi_{ph}$, are the thumb, index and middle phalanges respectively. According to human hand morphology, $mid$ and $dist$ subscripts refer to the middle and distal phalanx (e.g. $thumb_{dist}$ references the distal phalanx of thumb finger and at the implementation level it is a boolean value).

\subsection{Implementation details}

Grasping system has been originally implemented in Unreal Engine 4 (UE4), however, it can be easily implemented in other engines such as Unity, which would also provide us with the necessary components for replicating the system (e.g. overlapping triggers). The implementation consists of UE4 blueprints and has been correctly structured in the components depicted in Figure \ref{fig:pipeline} and described in the previous section. Implementation is available at Github\footnote{\url{https://github.com/3dperceptionlab/unrealgrasp}}.

\section{Performance analysis}
\label{sec:experimentation}

In order to validate our proposal, a complete performance analysis has been carried out. This analysis covers from a qualitative evaluation, which is prevalent in the assessment of VR systems, to a novel quantitative evaluation. Evaluation methods are briefly described as follows:
\begin{itemize}
	\item Qualitative evaluation: based on the user experience interacting with real objects from the YCB dataset in a photorealistic indoor scenario. Its purpose is to assess interaction realism, immersion, hand movement naturalness and other qualitative aspects described in Table \ref{table:questionnaire} from the Subsection \ref{subsec:qualitative}, which addresses qualitative evaluation in detail.
	\item Quantitative evaluation: based on the grasping quality in terms of realism (i.e. how much it is visually plausible). We consider a visually plausible grasp when hand palm or fingers are level with the object surface, as in a real life grasping. However, when dealing with complex meshes, the collision detection precision can be significantly influenced. In this case, fingers could penetrate the object surface, or remain above its surface when a collision was detected earlier than expected. This would result in an unnatural and unrealistic grasp. To visually quantify grasping quality, we purpose a novel error metric based on computing the distance from each capsule trigger to the nearest contact point on the object surface. Quantitative evaluation and the proposed error metric are addressed in detail in Subsection \ref{subsec:quantitative}.
	
\end{itemize}

\subsection{Qualitative evaluation}
\label{subsec:qualitative}

Most VR experiments include qualitative and quantitative studies to measure its realism and immersion. Arguably, questionnaires are the default method to qualitatively assess any experience and the vast majority of works include them in one way or another \cite{Christopoulos2018} \cite{Koutsabasis2018} \cite{Vosinakis2018}. However, one of the main problems with them is the absence of a standardized set of questions for different experiences that allows for fair and easy comparisons. The different nature of the VR systems and experiences makes it challenging to find a set of evaluation questions that fits them all. Following the efforts of \cite{Gonzalez-Franco2018} towards a standardized embodiment questionnaire, we analyzed several works in the literature \cite{Poeschl2013} \cite{Brackney2017} that included questionnaires to assess VR experiences to devise a standard one for virtual grasping systems. Inspired by such works, we have identified three main types of questions or aspects:
\begin{itemize}
	\item Motor Control: this aspect considers the movement of the virtual hands as a whole and its responsiveness to the virtual reality controllers. Hands should move naturally and their movements must be caused exactly by the controllers without unwanted movements and without limiting or restricting real movements to adapt to the virtual ones.
	\item Finger Movement: this aspect takes the specific finger movement into account. Such movements must be natural and plausibly. Moreover, they must react properly to the user's intent.
	\item Interaction Realism: this aspect is related to the interaction of the hand and fingers with objects. 
\end{itemize}

The questionnaire, shown in Table \ref{table:questionnaire}, is composed of fourteen questions related to the previously described aspects.
\begin{table}[!htb]
	\resizebox{\linewidth}{!}{
		\begin{tabular}{cl}
			\hline
			\textbf{ID} & \textbf{Question}\\
			\hline
			\multicolumn{2}{c}{\emph{Aspect 1: Motor Control}}\\
			Q1 & I felt like I could control the virtual hands as if it were my own hands\\
			Q2 & The movements of the virtual hands were caused by my movements\\
			Q3 & I felt as if the movements of the virtual hands were influencing my own movements\\
			Q4 & I felt as if the virtual hands were moving by themselves\\
			\hline
			\multicolumn{2}{c}{\emph{Aspect 2: Finger Movement Realism}}\\
			\hline
			Q5 & It seemed that finger movements were smooth and plausible\\
			Q6 & I felt fingers open and close in a natural way\\
			Q7 & Fingers react adequately to my intentions\\
			\hline
			\multicolumn{2}{c}{\emph{Aspect 3: Interaction Realism}}\\
			\hline
			Q8 & I felt like I could grab objects wherever I wanted to\\
			Q9 & It seemed as if the virtual fingers were mine when grabbing an object\\
			Q10 & I felt that grabbing objects was clumsy and hard to achieve\\
			Q11 & It seemed as if finger movement were guided and unnatural\\
			Q12 & I felt that grasps were visually correct and natural\\
			Q13 & I felt that grasps were physically correct and natural\\
			Q14 & It seemed that fingers were adapting properly to the different geometries\\
			\hline
	\end{tabular}}
	\caption{User evaluation questionnaire.}
	\label{table:questionnaire}
\end{table}
Following \cite{Gonzalez-Franco2018}, the users of the study will be presented with such questions right after the end of the experience in a randomized order to limit context effects. In addition, questions must be answered following the 7-point Likert-scale: (+3) strongly agree, (+2) agree, (+1) somewhat agree, (0) neutral, (-1) somewhat disagree, (-2) disagree, and (-3) strongly disagree. Results will be presented as a single embodiment score using the following equations:
\begin{equation}\label{eq:6}
\begin{split}
\text{Motor Control} &= ((Q1 + Q2) - (Q3 + Q4)) / 4 \\
\text{Finger Movement Realism} &= (Q5 + Q6 + Q7) / 3 \\
\text{Interaction Realism} &= ((Q8 + Q9) - (Q10 + Q11) \\ 
& + Q12 + Q13 + Q14) / 7  
\end{split}
\end{equation}, using the results of each individual aspect, we obtain the total embodiment score as follows:
\begin{equation} \label{eq:5}
\begin{split}
\text{Score} &= (\text{Motor Control} + \text{Finger Movement Realism} \\
& + \text{Interaction Realism} * 2) / 4
\end{split}
\end{equation}
The interaction realism is the key aspect of this qualitative evaluation. So that, in the Equation \ref{eq:5} we emphasize this aspect by weighting it higher.  

\subsection{Quantitative evaluation}
\label{subsec:quantitative}

With the quantitative evaluation, we aim to evaluate grasping quality in terms of how much it is visually plausible or realistic. In other words, our purpose is to visually quantify our grasping performance, analyzing each finger position and how it fits the object mesh. When a collision is detected by a capsule trigger, we proceed with the calculation of the nearest distance between the finger phalanx surface (delimited by the capsule trigger) and the object mesh (see Equation \ref{eq:1}).

\begin{figure}
	\centering
	\includegraphics[width=0.75\linewidth]{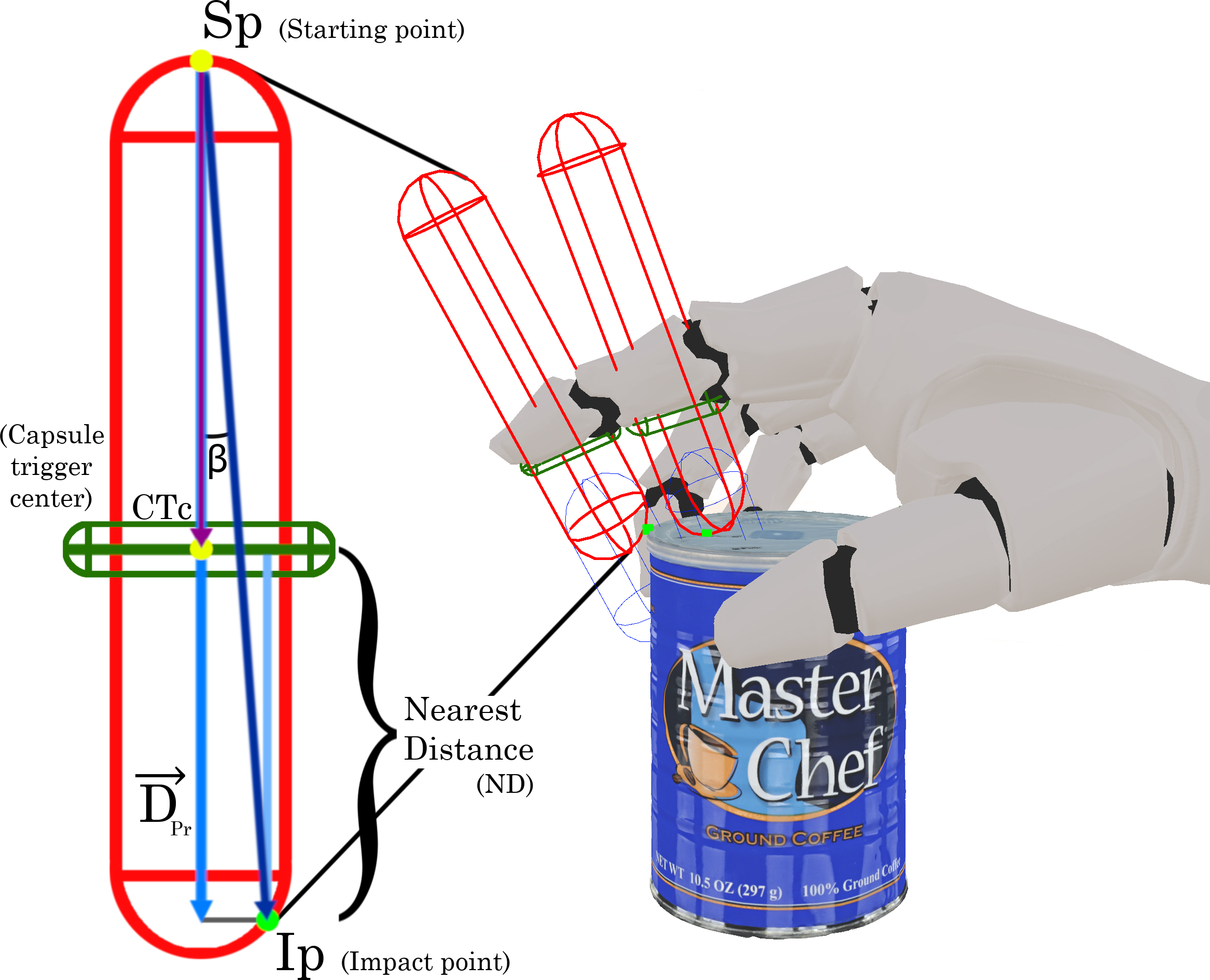}
	\caption{Distance computation for index finger.}
	\label{fig:distanceCalculus}
\end{figure}

In Figure \ref{fig:distanceCalculus} the red capsules are representing 3D sphere tracing volumes which provide information of the nearest collision from the trace starting point to the first contact point on the object surface which intersects the sphere volume. For each finger phalanx with an attached capsule trigger represented in green, we throw a sphere trace obtaining the nearest contact points on the object surface represented as lime colored dots (impact point, Ip). In this representation, the total error for the index finger would be the average of the sum of the distances in millimeters between the surface of each phalanx and the nearest contact point on the object surface (see Equation \ref{eq:2}).
The nearest distance computation is approximated by an equation that was developed to find the distance between the impact point, and the plane that contains the capsule trigger center point and is perpendicular to the longitudinal axis of the red capsule. Capsule triggers centers are located on the surface of the hand mesh, so this computation should approximate the nearest distance to the mesh well enough, without being computationally too demanding. To compute this distance, we define the following vectors from the three input points (the starting point of the red capsule, the impact point and the capsule trigger center point):
\begin{equation} \label{eq:1aux1}
\begin{split}
\overrightarrow{D_{Ip}} &= Ip - Sp\\
\overrightarrow{D_{CTc}} &= CTc - Sp
\end{split}
\end{equation}  
where $\overrightarrow{D_{Ip}}$ is the vector from the starting point to the impact point, and $\overrightarrow{D_{CTc}}$ vector represents the direction of the longitudinal axis of the red capsule. They are represented in navy blue and purple respectively in Figure \ref{fig:distanceCalculus}. Then, we find the cosine of the angle they form through their dot product:
\begin{equation} \label{eq:1aux2}
\begin{split}
\overrightarrow{D_{Ip}} \cdot \overrightarrow{D_{CTc}} &= |\overrightarrow{D_{Ip}}| * |\overrightarrow{D_{CTc}}| * cos(\beta)\\
cos(\beta) &= \frac{\overrightarrow{D_{Ip}} \cdot \overrightarrow{D_{CTc}}}{|\overrightarrow{D_{Ip}}| * |\overrightarrow{D_{CTc}}|}\\
\end{split}
\end{equation}
We can now substitute that cosine when computing the projection of $\overrightarrow{D_{Ip}}$ over the longitudinal axis of the red capsule ($\overrightarrow{D_{Pr}}$ in Figure \ref{fig:distanceCalculus}):
\begin{equation} \label{eq:1aux3}
\begin{split}
|\overrightarrow{D_{Pr}}| &= cos(\beta) * |\overrightarrow{D_{Ip}}|\\
|\overrightarrow{D_{Pr}}| &= \frac{\overrightarrow{D_{Ip}} \cdot \overrightarrow{D_{CTc}}}{|\overrightarrow{D_{CTc}}| * |\overrightarrow{D_{Ip}}|} * |\overrightarrow{D_{Ip}}|\\
|\overrightarrow{D_{Pr}}| &= \frac{\overrightarrow{D_{Ip}} \cdot \overrightarrow{D_{CTc}}}{|\overrightarrow{D_{CTc}}|}
\end{split}
\end{equation}
Having that module, we only have to subtract $|\overrightarrow{D_{CTc}}|$ in order to obtain the desired distance:  
\begin{equation} \label{eq:1}
\begin{split}
ND(Ip,Sp,CTc) &= \frac{\overrightarrow{D_{Ip}} \cdot \overrightarrow{D_{CTc}} }{|\overrightarrow{D_{CTc}}|} - |\overrightarrow{D_{CTc}}|\\
ND(Ip,Sp,CTc) &= \frac{\overrightarrow{Ip-Sp} \cdot \overrightarrow{CTc-Sp} }{|\overrightarrow{CTc-Sp}|} - |\overrightarrow{CTc-Sp}|
\end{split}
\end{equation}
Computing the distance like this, with this final subtraction, allows to obtain a positive distance when impact point is outside the hand mesh, and a negative one if it is inside.

We compute the nearest distance per each capsule trigger attached to a finger phalanx. As stated before, if the distance is negative, this indicates a finger penetration issue on the object surface. Otherwise, if distance is positive, it means that finger stopped above the object surface. The ideal case is when a zero distance is obtained, that is, the finger is perfectly situated on the object surface.

The total error for the hand is represented by the following equation:
\begin{equation} \label{eq:2}
HandError = \sum_{i=1}^{N_{Fingers}}\frac{\sum_{j=1}^{N_{CTF}}|ND(Ip_{ij},Sp_{ij},CTc_{ij})|}{N_{CapsuleTriggersPerFinger}}
\end{equation}

\subsection{Dataset}

To benchmark our grasping system we used a set of objects that are frequently used in daily life, such as, food items (e.g. cracker box, cans, box of sugar, fruits, etc.), tool items (e.g. power drill, hammer, screwdrivers, etc.), kitchen items (e.g. eating utensils) and also spherical shaped objects (e.g. tennis ball, racquetball, golf ball, etc.). Yale-CMU-Berkeley (YCB) Object and Model set \cite{calli2015} provides us these real-life 3D textured models scanned with outstanding accuracy and detail. Available objects have a wide variety of shapes, textures and sizes as we can see in Figure \ref{fig:ycbgrasps}. The advantage of using real life objects is that the user already has a previous experience manipulating similar objects so he will try to grab and interact with the objects in the same way.

\subsection{Participants}

For the performance analysis, we recruited ten participants (8M/2F) from the local campus. Four of them have experience with VR applications. The rest are inexperienced virtual reality users. Participants will take part on both qualitative and quantitative evaluation. The performance analysis procedure will be described in the following subsection, indicating the concrete tasks to be performed by each participant.

\subsection{Procedure}

\begin{figure*}[!htb]
	\centering
	\begin{subfigure}[t]{0.15\textwidth}
		\includegraphics[trim={9cm 1cm 8cm 6.3cm},clip,width=\textwidth]{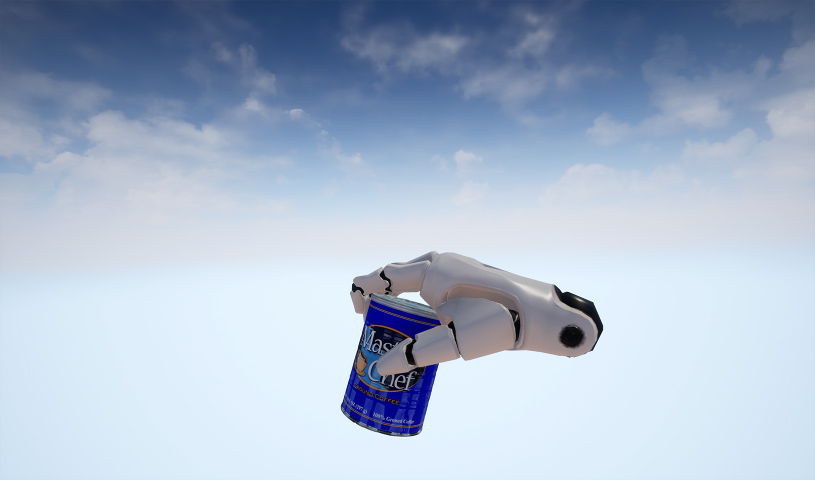}
		\caption{Master Chef can}
		\label{fig:masterchefcan}
	\end{subfigure}
	\begin{subfigure}[t]{0.15\textwidth}
		\includegraphics[trim={8cm 0cm 7cm 5cm},clip,width=\textwidth]{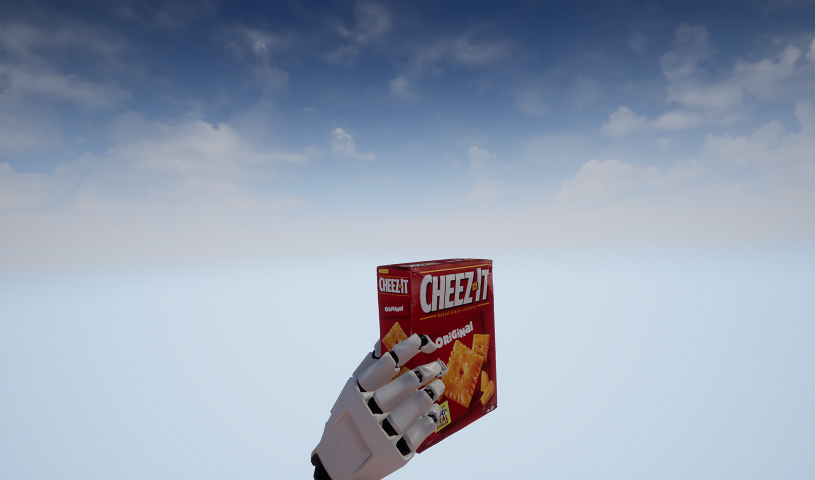}
		\caption{Cracker box}
		\label{fig:crackerbox}
	\end{subfigure}
	\begin{subfigure}[t]{0.15\textwidth}
		\includegraphics[trim={9cm 1.5cm 8.5cm 6.5cm},clip, width=\textwidth]{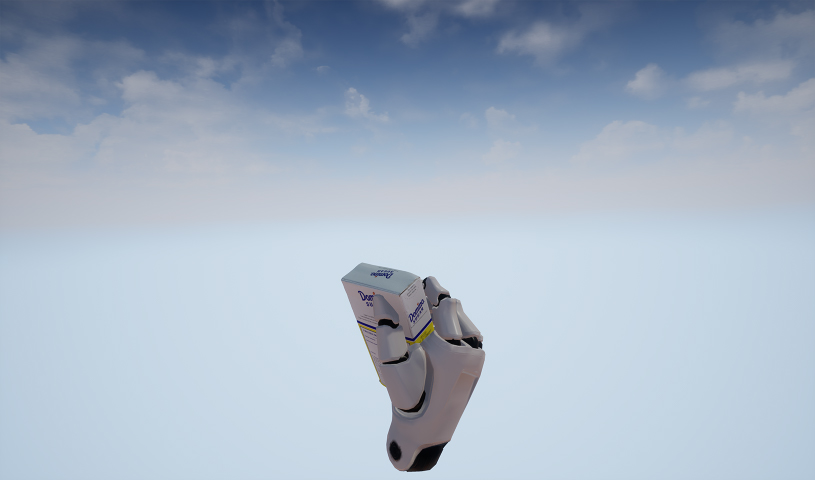}
		\caption{Sugar box}
		\label{fig:sugarbox}
	\end{subfigure}
	\begin{subfigure}[t]{0.15\textwidth}
		\includegraphics[trim={9.2cm 1.5cm 8.7cm 7cm},clip, width=\textwidth]{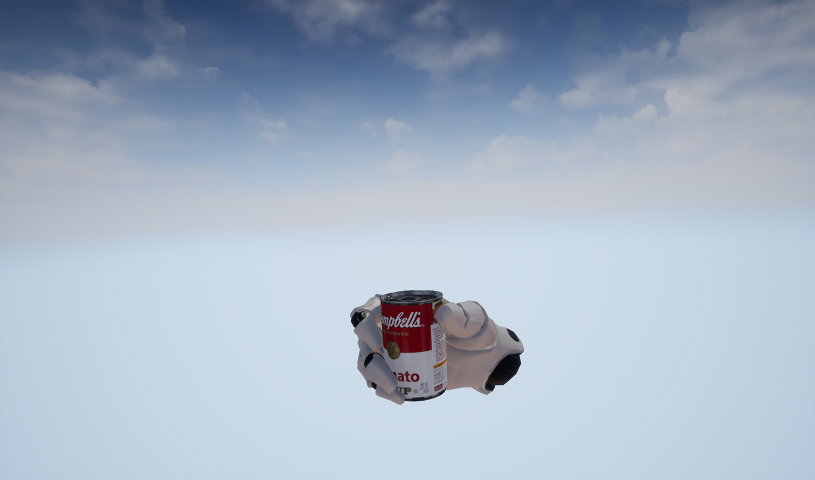}
		\caption{Tomato soup can}
		\label{fig:tomatosoupcan}
	\end{subfigure}
	\begin{subfigure}[t]{0.15\textwidth}
		\includegraphics[trim={8cm 1cm 8cm 5.5cm},clip, width=\textwidth]{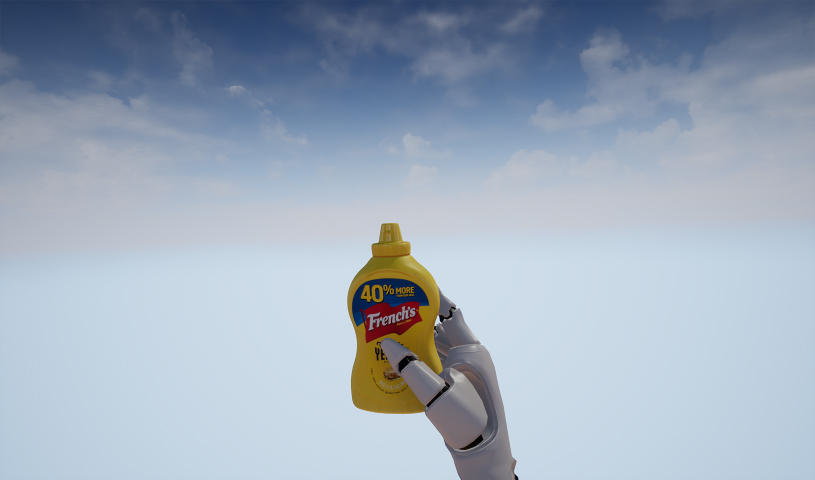}
		\caption{Mustard bottle}
		\label{fig:mustardbottle}
	\end{subfigure}
	\begin{subfigure}[t]{0.15\textwidth}
		\includegraphics[trim={9cm 1.5cm 7cm 5cm},clip, width=\textwidth]{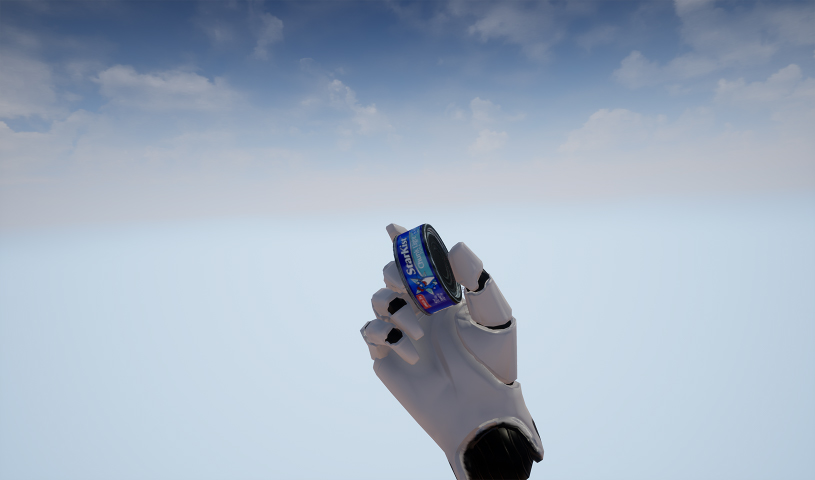}
		\caption{Tuna fish can}
		\label{fig:tunafishcan}
	\end{subfigure}
	\begin{subfigure}[t]{0.15\textwidth}
		\includegraphics[trim={9.5cm 1.5cm 7cm 5.5cm},clip, width=\textwidth]{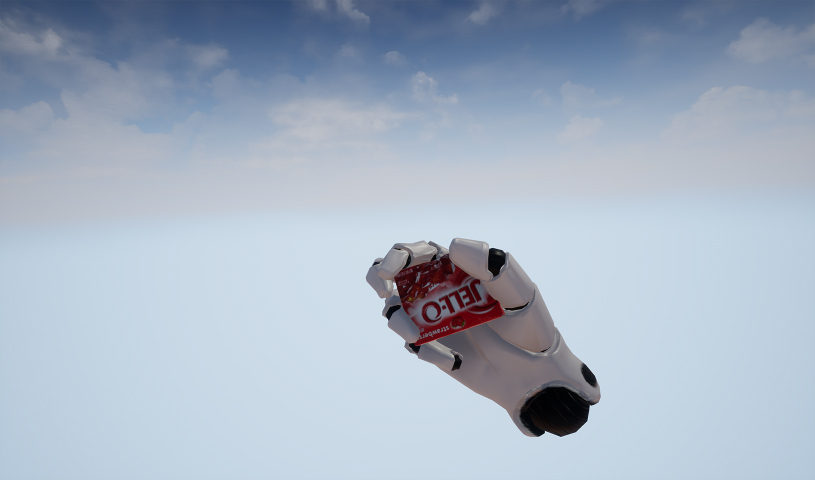}
		\caption{Gelatin box}
		\label{fig:gelatinbox}
	\end{subfigure}
	\centering
	\begin{subfigure}[t]{0.15\textwidth}
		\includegraphics[trim={8.5cm 0cm 7.5cm 7cm},clip, width=\textwidth]{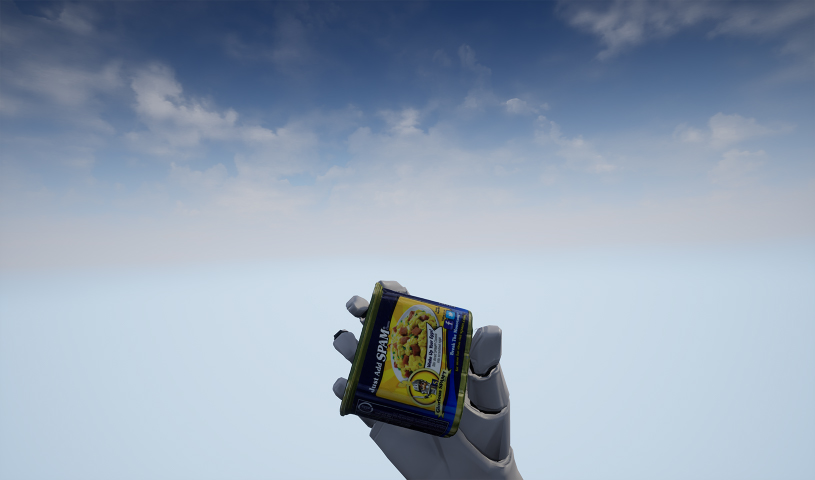}
		\caption{Potted meat can}
		\label{fig:pottedmeatcan}
	\end{subfigure}
	\begin{subfigure}[t]{0.15\textwidth}
		\includegraphics[trim={9cm 1cm 6cm 5cm},clip, width=\textwidth]{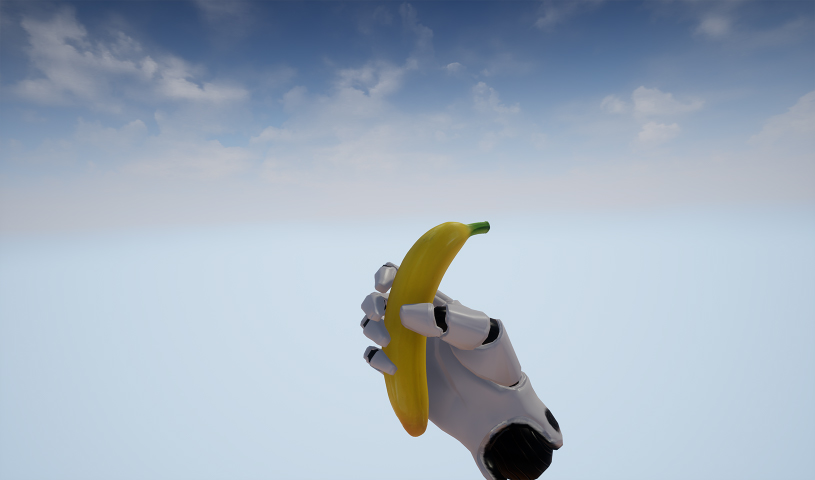}
		\caption{Banana}
		\label{fig:banana}
	\end{subfigure}
	\begin{subfigure}[t]{0.15\textwidth}
		\includegraphics[trim={10cm 2.5cm 7cm 5.5cm},clip, width=\textwidth]{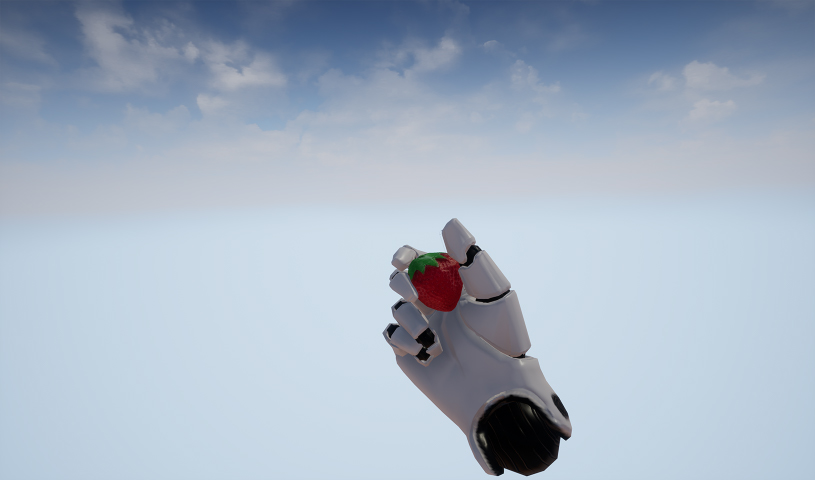}
		\caption{Strawberry}
		\label{fig:strawberry}
	\end{subfigure}
	\begin{subfigure}[t]{0.15\textwidth}
		\includegraphics[trim={9.5cm 1.5cm 7cm 6cm},clip, width=\textwidth]{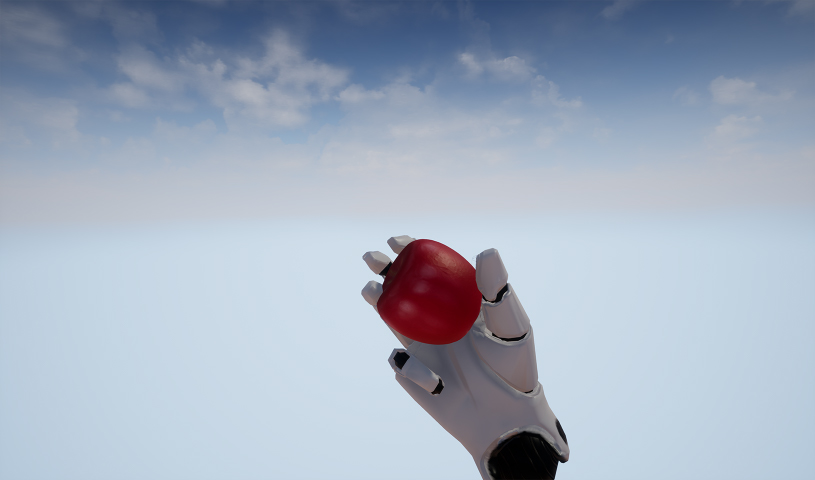}
		\caption{Apple}
		\label{fig:apple}
	\end{subfigure}
	\begin{subfigure}[t]{0.15\textwidth}
		\includegraphics[trim={10cm 1.5cm 7cm 6.5cm},clip, width=\textwidth]{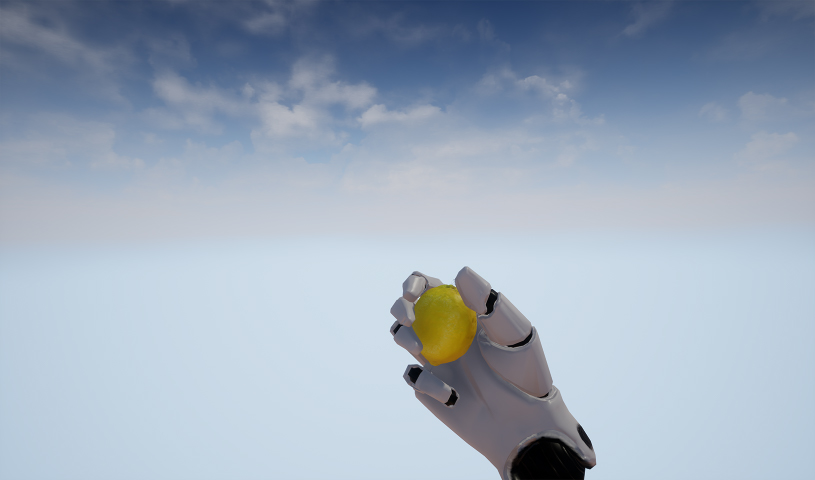}
		\caption{Lemon}
		\label{fig:lemon}
	\end{subfigure}
	\begin{subfigure}[t]{0.15\textwidth}
		\includegraphics[trim={9.5cm 2.08cm 8cm 6cm},clip, width=\textwidth]{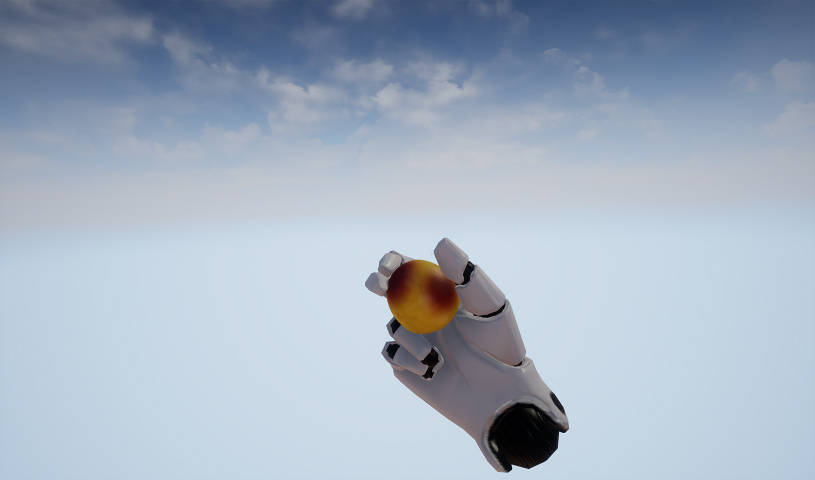}
		\caption{Peach}
		\label{fig:peach}
	\end{subfigure}
	\begin{subfigure}[t]{0.15\textwidth}
		\includegraphics[trim={9.5cm 2.3cm 7.5cm 5.2cm},clip, width=\textwidth]{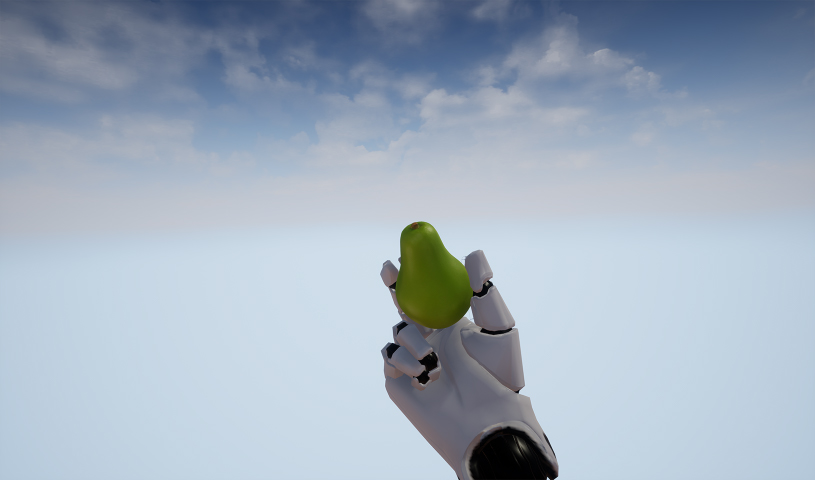}
		\caption{Pear}
		\label{fig:pear}
	\end{subfigure}
	\begin{subfigure}[t]{0.15\textwidth}
		\includegraphics[trim={9.7cm 2cm 8cm 6.3cm},clip, width=\textwidth]{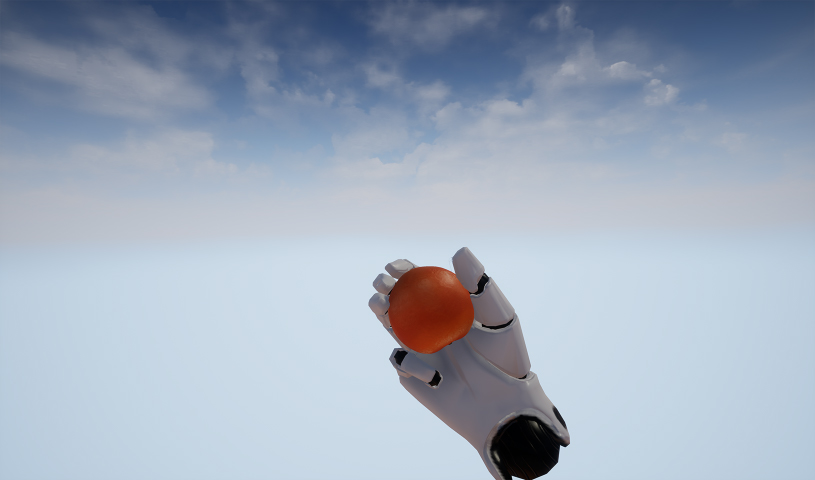}
		\caption{Orange}
		\label{fig:orange}
	\end{subfigure}
	\begin{subfigure}[t]{0.15\textwidth}
		\includegraphics[trim={9.5cm 2cm 7cm 5.5cm},clip, width=\textwidth]{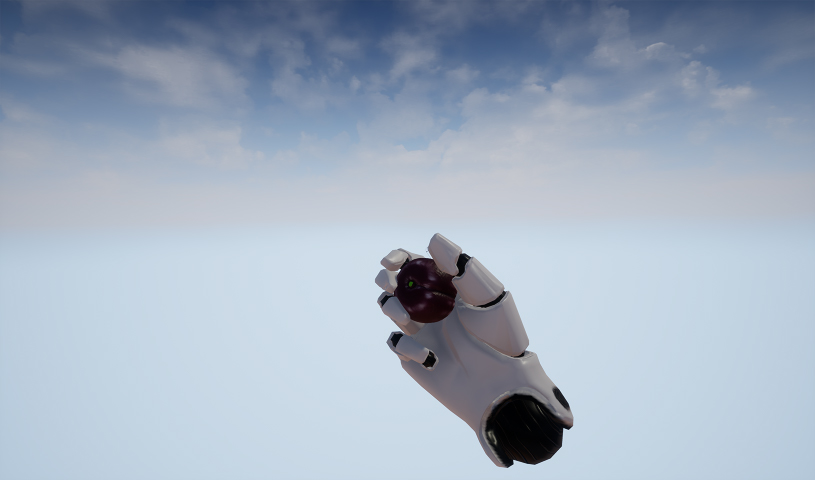}
		\caption{Plum}
		\label{fig:plum}
	\end{subfigure}
	\begin{subfigure}[t]{0.15\textwidth}
		\includegraphics[trim={7cm 1cm 7cm 4cm},clip, width=\textwidth]{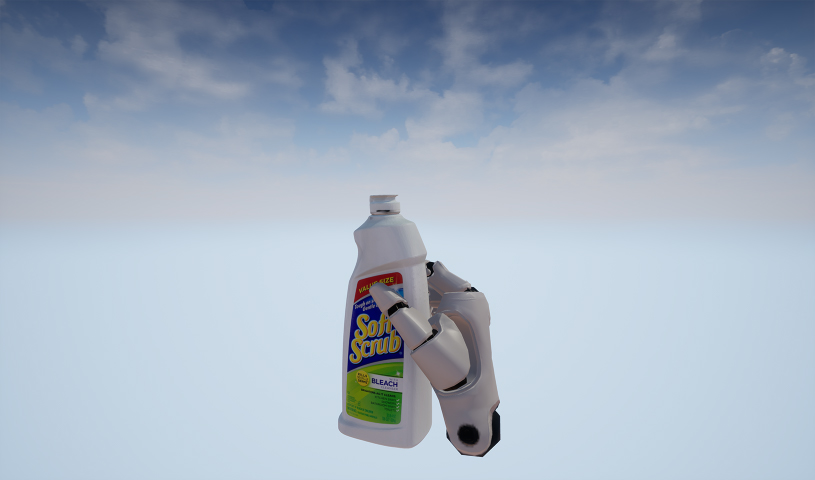}
		\caption{Bleach cleaner}
		\label{fig:bleachcleaner}
	\end{subfigure}
	\begin{subfigure}[t]{0.15\textwidth}
		\includegraphics[trim={8cm 2cm 6cm 5cm},clip, width=\textwidth]{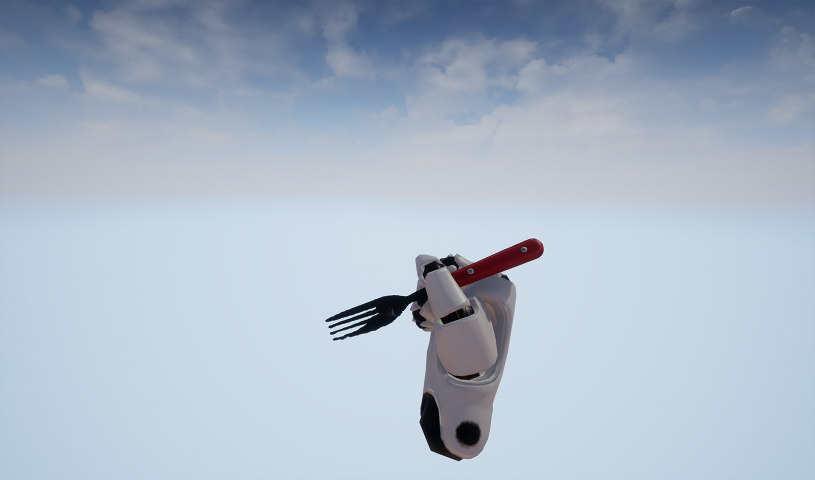}
		\caption{Fork}
		\label{fig:fork}
	\end{subfigure}
	\begin{subfigure}[t]{0.15\textwidth}
		\includegraphics[trim={9cm 1.8cm 6cm 6cm},clip, width=\textwidth]{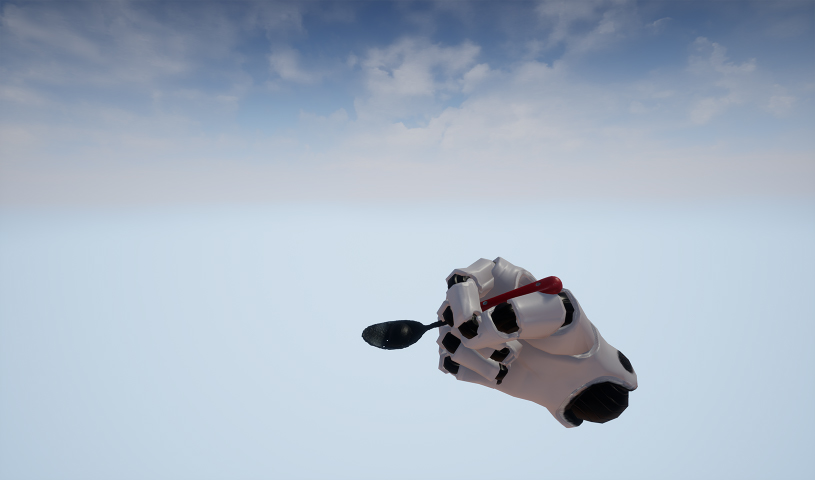}
		\caption{Spoon}
		\label{fig:spoon}
	\end{subfigure}
	\begin{subfigure}[t]{0.15\textwidth}
		\includegraphics[trim={9cm 4.2cm 6cm 3.5cm},clip, width=\textwidth]{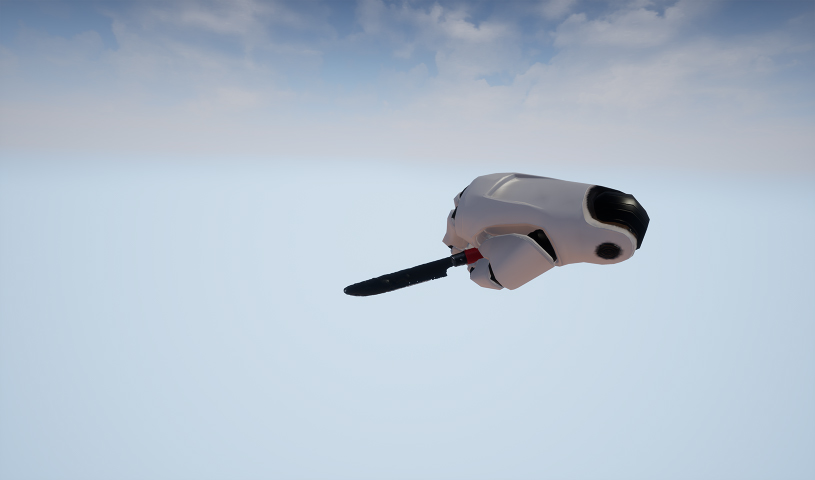}
		\caption{Knife}
		\label{fig:knife}
	\end{subfigure}
	\begin{subfigure}[t]{0.15\textwidth}
		\includegraphics[trim={8cm 2cm 6cm 5cm},clip, width=\textwidth]{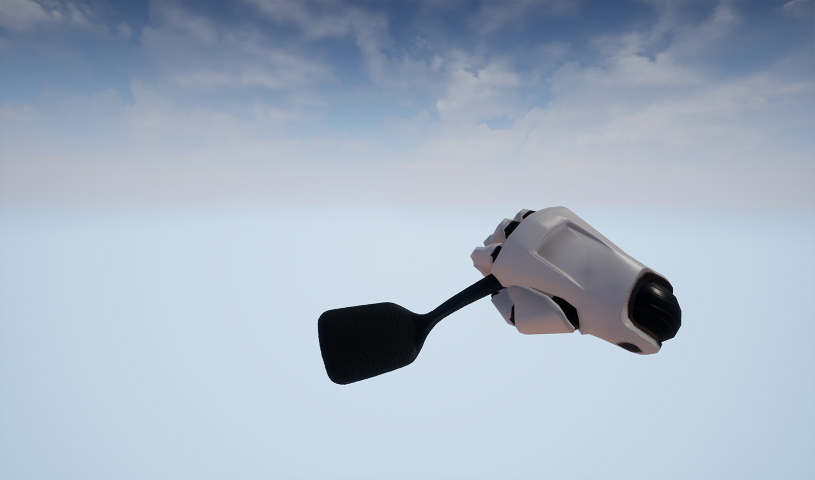}
		\caption{Spatula}
		\label{fig:spatula}
	\end{subfigure}
	\begin{subfigure}[t]{0.15\textwidth}
		\includegraphics[trim={7cm 1cm 7cm 5cm},clip, width=\textwidth]{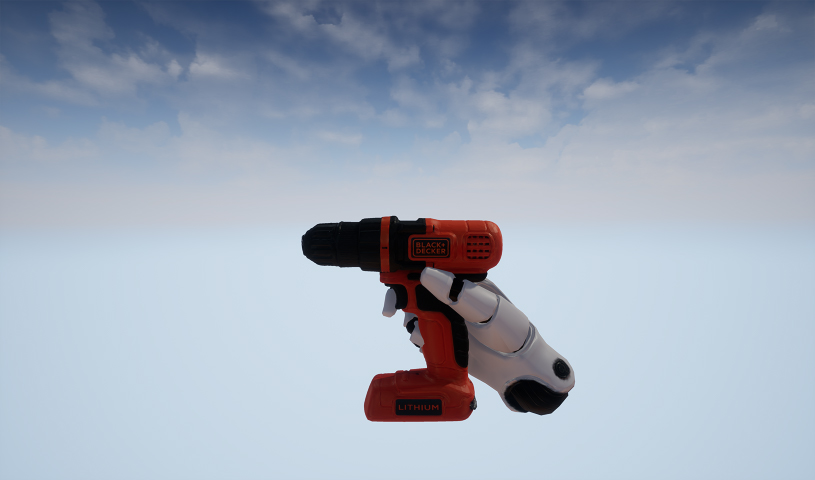}
		\caption{Power drill}
		\label{fig:powerdrill}
	\end{subfigure}
	\begin{subfigure}[t]{0.15\textwidth}
		\includegraphics[trim={8cm 3cm 7cm 5cm},clip, width=\textwidth]{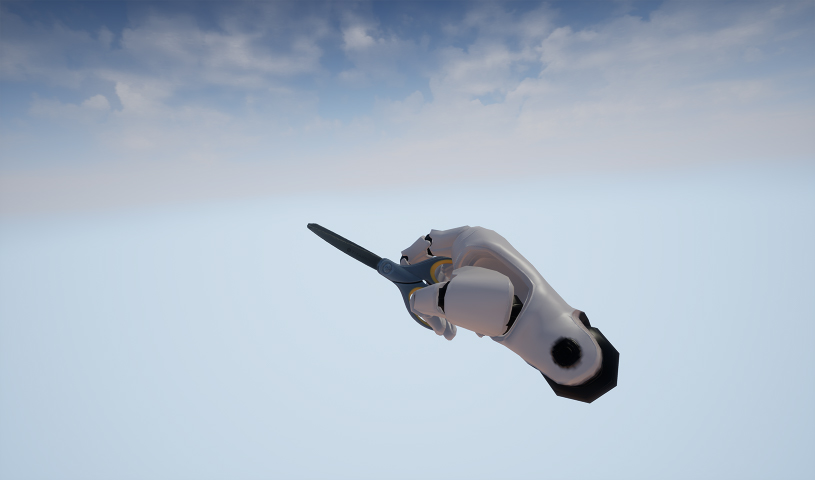}
		\caption{Scissors}
		\label{fig:scissors}
	\end{subfigure}
	\begin{subfigure}[t]{0.15\textwidth}
		\includegraphics[trim={10cm 3cm 8cm 5.5cm},clip, width=\textwidth]{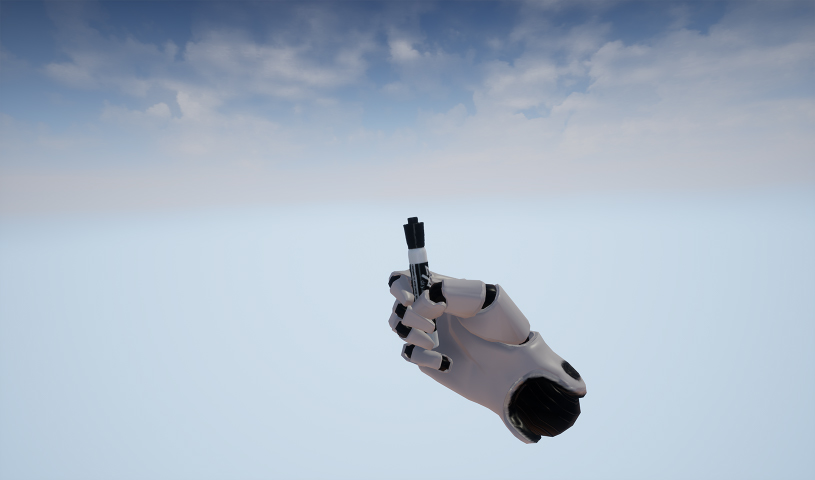}
		\caption{Large marker}
		\label{fig:largemarker}
	\end{subfigure}
	\begin{subfigure}[t]{0.15\textwidth}
		\includegraphics[trim={9cm 2cm 7cm 4cm},clip, width=\textwidth]{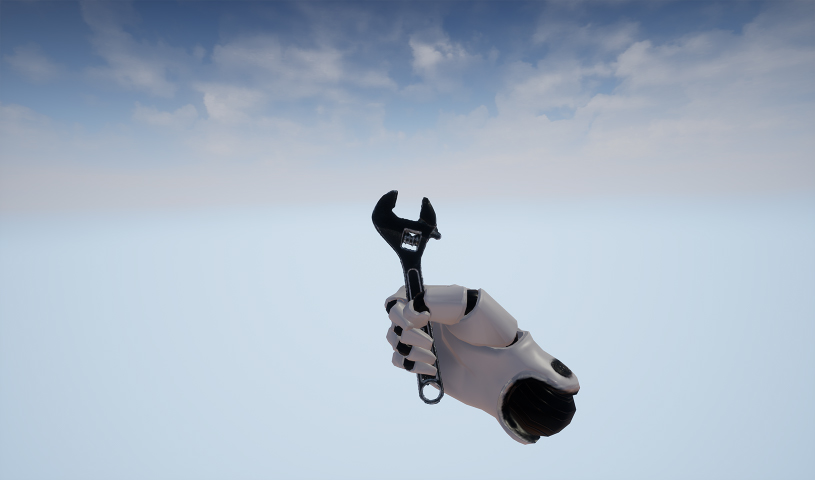}
		\caption{Adjustable wrench}
		\label{fig:adjustablewrench}
	\end{subfigure}
	\begin{subfigure}[t]{0.15\textwidth}
		\includegraphics[trim={7.5cm 2cm 8cm 4.5cm},clip, width=\textwidth]{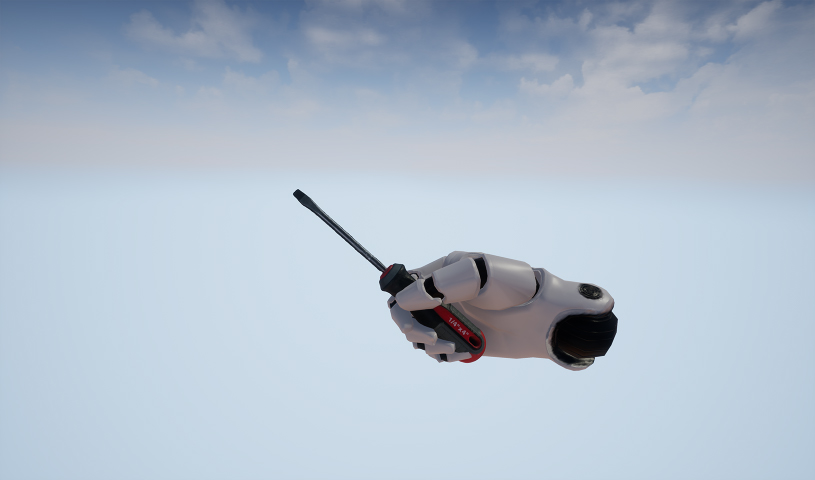}
		\caption{Flat screwdriver}
		\label{fig:flatscrewdriver}
	\end{subfigure}
	\begin{subfigure}[t]{0.15\textwidth}
		\includegraphics[trim={7cm 0.8cm 6cm 3cm},clip, width=\textwidth]{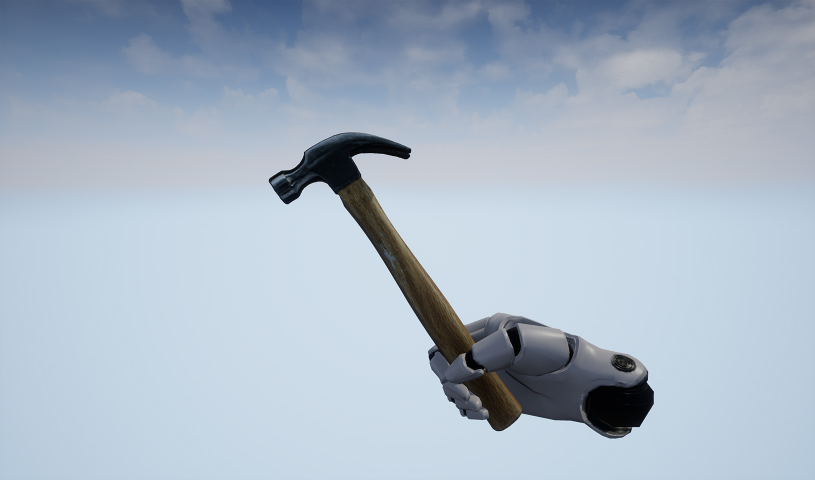}
		\caption{Hammer}
		\label{fig:hammer}
	\end{subfigure}
	\begin{subfigure}[t]{0.15\textwidth}
		\includegraphics[trim={10cm 2.5cm 7cm 5.5cm},clip, width=\textwidth]{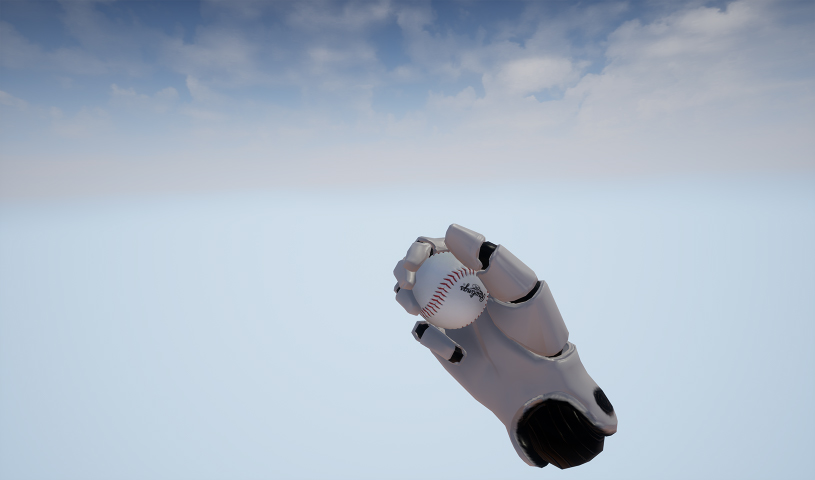}
		\caption{Baseball}
		\label{fig:baseball}
	\end{subfigure}
	\begin{subfigure}[t]{0.15\textwidth}
		\includegraphics[trim={9.5cm 2cm 7cm 5.5cm},clip, width=\textwidth]{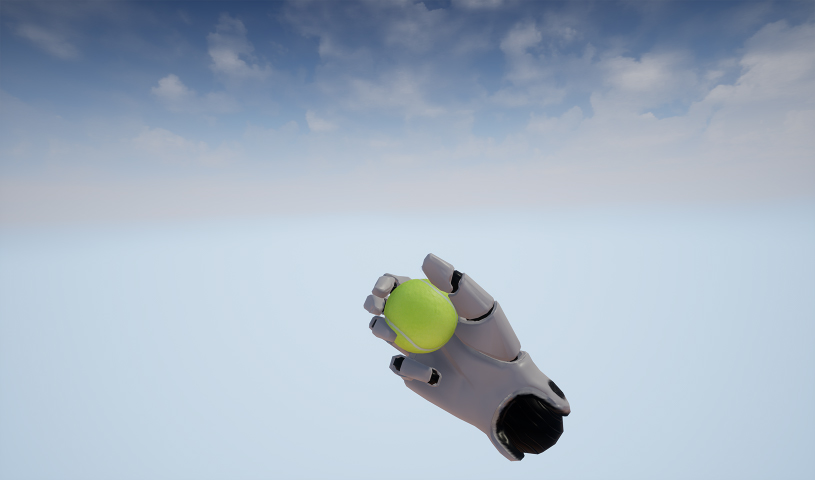}
		\caption{Tennis ball}
		\label{fig:tennisball}
	\end{subfigure}
	\begin{subfigure}[t]{0.15\textwidth}
		\includegraphics[trim={6cm 1.5cm 7cm 4cm},clip, width=\textwidth]{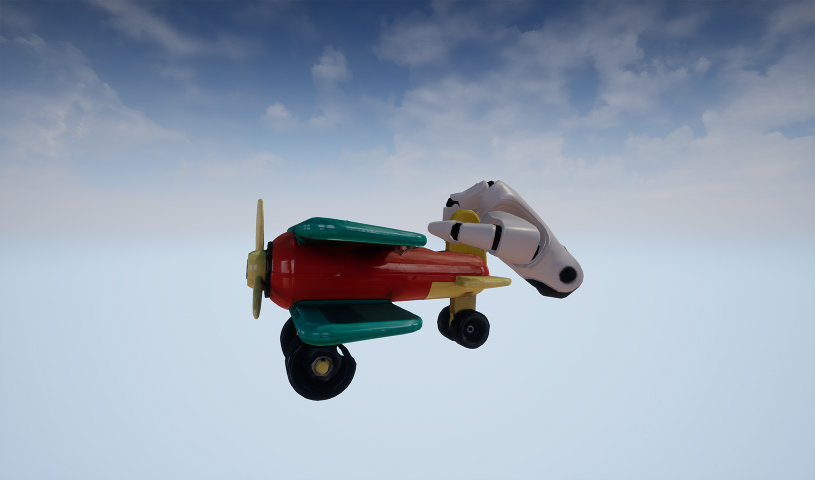}
		\caption{Toy airplane}
		\label{fig:toyairplane}
	\end{subfigure}
	\caption{Grasping performed on objects from the YCB dataset.}\label{fig:ycbgrasps}
\end{figure*}

The system performance analysis begins with the quantitative evaluation. In this first phase, the user will be embodied in a controlled scenario\footnote{\url{https://youtu.be/4sPhLbHpywM}} where 30 different objects will be spawned in a delimited area, with random orientation, and in the same order as represented in Figure \ref{fig:ycbgrasps}. The user will try to grasp the object as he would do in real life and as quickly as possible. For each grasping, the system will compute the error metric and will also store the time spent by the user in grasping the object. The purpose of this first phase is to visually analyze grasping quality which is directly related to user expertise in VR environments and concretely with our grasping system. An experienced user would know system limits both when interacting with complex geometries or with large objects that would make it difficult to perform the grasp action quickly and naturally.

For the qualitative evaluation, the same user will be embodied in a photorealistic scenario changing mannequin hands by human hand model with realistic textures. After interacting freely in the photorealistic virtual environment\footnote{\url{https://youtu.be/65gdFdwsTVg}}, the user will have to answer the evaluation questionnaire defined in Table \ref{table:questionnaire}. The main purpose is the evaluation of interaction realism, finger and hand movement naturalness and motor control, among other qualitative aspects regarding user experience in VR environments. 

\section{Results and discussion}
\label{sec:results}

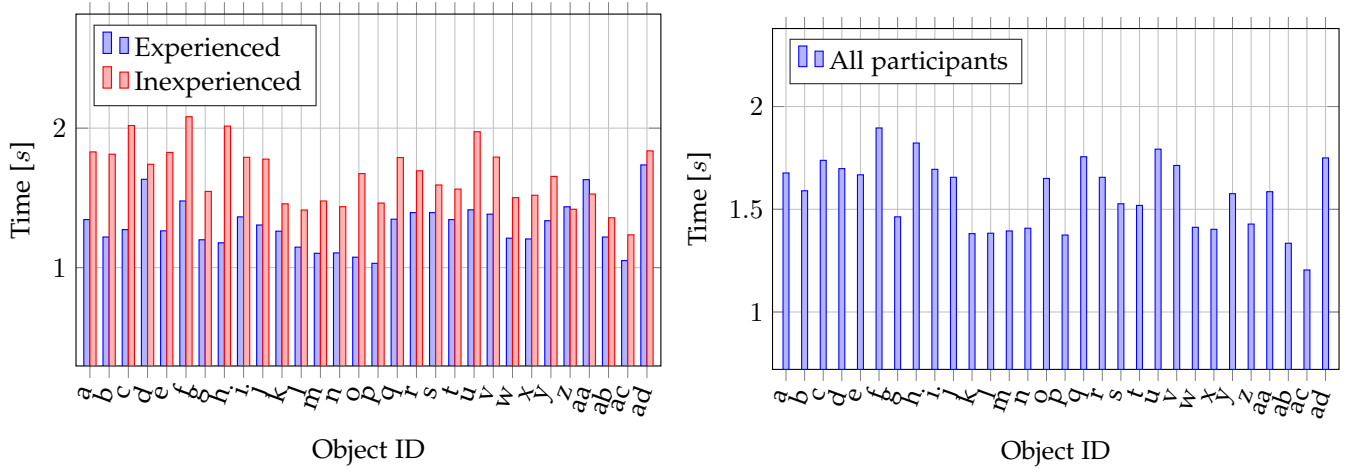
\begin{figure*}
	\centering
	\begin{subfigure}[b]{0.49\linewidth}
		\resizebox{\linewidth}{!}{
			\begin{tikzpicture}
			\begin{axis}[
			ybar= 0pt,
			xtick=data,
			symbolic x coords= {a,b,c,d,e,f,g,h,i,j,k,l,m,n,o,p,q,r,s,t,u,v,w,x,y,z,aa,ab,ac,ad},
			x tick label style={rotate=70, anchor=east, align=right},
			enlargelimits=false,
			enlarge x limits=0.025,
			enlarge y limits=0.7,
			bar width=0.23 em,
			grid=both,
			grid style={line width=.1pt, draw=gray!10},
			major grid style={line width=.2pt,draw=gray!50},
			ylabel={Time [$s$]},
			xlabel={Object ID},
			width=\linewidth,
			height=6cm,
			legend style={legend cell align=left, align=left, draw=white!15!black},
			legend pos=north west
			]
			\addplot table[x=object,y=0]{Data/time_per_object.dat};
			\addplot table[x=object,y=1]{Data/time_per_object.dat};
			\legend{Experienced, Inexperienced}
			\end{axis}
			\end{tikzpicture}}
	\end{subfigure}
	\begin{subfigure}[b]{0.49\linewidth}
		\centering
		\resizebox{\linewidth}{!}{
			\begin{tikzpicture}
			\begin{axis}[
			ybar= 0pt,
			xtick=data,
			symbolic x coords= {a,b,c,d,e,f,g,h,i,j,k,l,m,n,o,p,q,r,s,t,u,v,w,x,y,z,aa,ab,ac,ad},
			x tick label style={rotate=70, anchor=east, align=center},
			enlargelimits=false,
			enlarge x limits=0.025,
			enlarge y limits=0.7,
			bar width=0.23 em,
			grid=both,
			grid style={line width=.1pt, draw=gray!10},
			major grid style={line width=.2pt,draw=gray!50},
			ylabel={Time [$s$]},
			xlabel={Object ID},
			width=\linewidth,
			height=6cm,
			legend style={legend cell align=left, align=left, draw=white!15!black},
			legend pos=north west
			]
			\addplot table[x=object,y=0]{Data/time_per_object_overall.dat};
			\legend{All participants}
			\end{axis}
			\end{tikzpicture}}
	\end{subfigure}
	\caption{At left, average elapsed time obtained by each participants group on grasping each object. At right, average elapsed time obtained by all the participants on grasping each object.}
	\label{chart:time}
\end{figure*}

In this section we will discuss and analyze the results obtained from the performance analysis process. On the one hand, we will draw conclusions from the average error obtained in grasping each object by each participant group, and also from the overall error per object taking into account all the participants (see Figure \ref{chart:error}). On the other hand, we obtained the average elapsed time needed in grasping each object for each participant group, and also the average elapsed time needed for each object taking into account all the participants (see Figure \ref{chart:time}). This will allow us to draw conclusions about the most difficult objects to manipulate in terms of accuracy and elapsed time for grasping. Moreover, we can compare system performance used by inexperienced users in comparison with experienced ones.

\subsection{Qualitative evaluation}

Qualitative evaluation for each participant was calculated using the Equation \ref{eq:6} obtaining a score for each qualitative aspect. In Table \ref{table:qualitativeresults} we represent for each group of participants: the average score for each evaluation aspect and the total embodiment score computed using the Equation \ref{eq:5}. 
\begin{table}[!htb]
	\resizebox{\linewidth}{!}{
		\begin{tabular}{lcc}
			\hline
			& \multicolumn{2}{c}{Score} \\
			\cline{2-3}
			Evaluation Aspects    & Experienced users & Inexperienced users \\
			\hline
			(1) Motor Control     & 1.85    & 2.34      \\
			(2) Finger Movement Realism & 2.33   & 2.51       \\
			(3) Interaction Realism      & 1.84     & 1.95      \\
			\hline
			Embodiment score & 1.97 &  2.19 \\
			\hline
	\end{tabular}}
	\caption{Average score for each qualitative aspect of the evaluation and group of participants. Maximum result would be three.}
	\label{table:qualitativeresults}
\end{table}
Regarding the represented results in Table \ref{table:qualitativeresults}, the evaluation of experienced users has been more disadvantageous as they have a more elaborated criterion given their previous experience with virtual reality applications. Finger movement realism (aspect 2) was evaluated similarly by both groups. This is because the hand closing and opening gestures are guided by the same animation in both cases. Finally, the reported results referring to the interaction realism have been the lowest in both cases. This is mostly because users cannot control their individual fingers movement, since general hand gesture is controlled by a unique trigger button of the controller. However, overall obtained embodiment score is 2.08 out of 3.0.

\subsection{Quantitative evaluation}

As expected, inexperienced users have taken longer to grasp almost all the object set due to the lack of practice and expertise with the system. This is clearly represented in Figure \ref{chart:time} where experienced users only have taken longer in grasping some tools such as, the flat screwdriver (Figure \ref{fig:flatscrewdriver}) and hammer (Figure \ref{fig:hammer}). Inexperienced users take an average of 0.36 seconds longer to grab the objects. In practice and regarding interaction, this is not a factor that is going to make a crucial difference. Analyzing Figure \ref{chart:time}, the tuna fish can (Figure \ref{fig:tunafishcan}), potted meat can (Figure \ref{fig:pottedmeatcan}), spatula (Figure \ref{fig:spatula}), toy airplane (Figure \ref{fig:toyairplane}) and bleach cleaner (Figure \ref{fig:bleachcleaner}) are the most time consuming when grasped by the users. This is mainly because of their sizes and complex geometries. Since objects are spawned with a random orientation, this fact can affect grasping times. Even so, we can conclude that the largest objects are those that the user takes the longest to grasp.

\begin{figure*}
	\centering
	\begin{subfigure}[b]{0.49\textwidth}
		\resizebox{\linewidth}{!}{
			\begin{tikzpicture}
			\begin{axis}[
			ybar= 0pt,
			xtick=data,
			symbolic x coords= {a,b,c,d,e,f,g,h,i,j,k,l,m,n,o,p,q,r,s,t,u,v,w,x,y,z,aa,ab,ac,ad},
			x tick label style={rotate=70, anchor=east, align=right},
			enlargelimits=false,
			enlarge x limits=0.025,
			enlarge y limits=0.05,
			bar width=0.23 em,
			grid=both,
			grid style={line width=.1pt, draw=gray!10},
			major grid style={line width=.2pt,draw=gray!50},
			ylabel={Error [$mm$]},
			xlabel={Object ID},
			width=\linewidth,
			height=6cm,
			legend style={legend cell align=left, align=left, draw=white!15!black},
			legend pos=north west
			]
			\addplot table[x=object,y=0]{Data/accuracy_per_object.dat};
			\addplot table[x=object,y=1]{Data/accuracy_per_object.dat};
			\legend{Experienced, Inexperienced}
			\end{axis}
			\end{tikzpicture}}
	\end{subfigure}
	\begin{subfigure}[b]{0.49\textwidth}
		\resizebox{\linewidth}{!}{
			\begin{tikzpicture}
			\begin{axis}[
			ybar= 0pt,
			xtick=data,
			symbolic x coords= {a,b,c,d,e,f,g,h,i,j,k,l,m,n,o,p,q,r,s,t,u,v,w,x,y,z,aa,ab,ac,ad},
			x tick label style={rotate=70, anchor=east, align=center},
			enlargelimits=false,
			enlarge x limits=0.025,
			enlarge y limits=0.05,
			bar width=0.23 em,
			grid=both,
			grid style={line width=.1pt, draw=gray!10},
			major grid style={line width=.2pt,draw=gray!50},
			ylabel={Error [$mm$]},
			xlabel={Object ID},
			width=\linewidth,
			height=5cm,
			legend style={legend cell align=left, align=left, draw=white!15!black},
			legend pos=north west
			]
			\addplot table[x=object,y=0]{Data/accuracy_per_object_overall.dat};
			\legend{All participants}
			\end{axis}
			\end{tikzpicture}}
	\end{subfigure}
	\caption{At left, average error(mm) obtained by each participants group and for each object.At right, average error(mm) obtained by all the participants and for each object.}
	\label{chart:error}
\end{figure*}
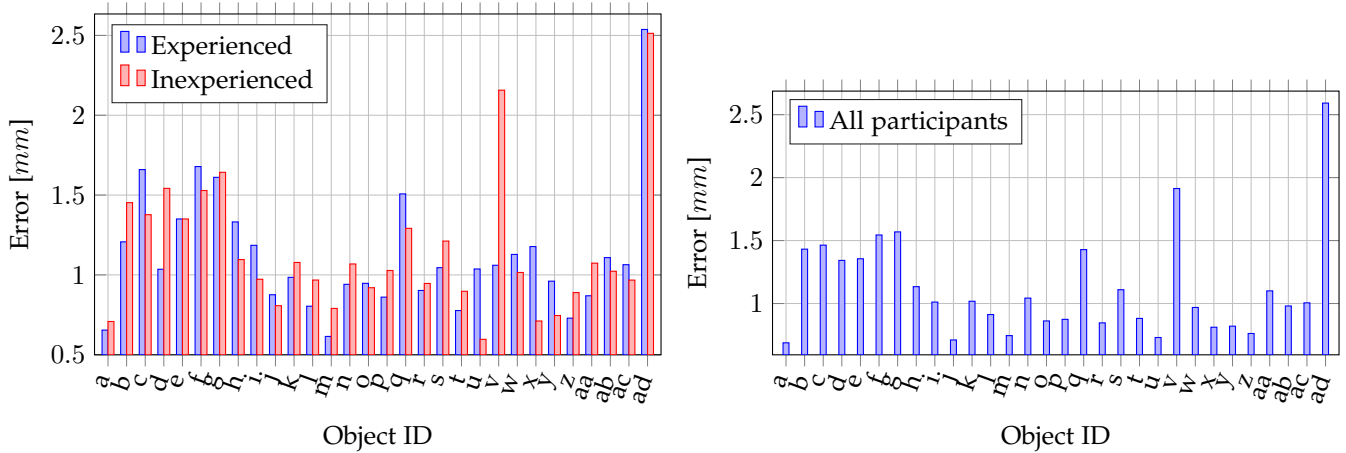

Regarding Figure \ref{chart:error} we can observe that the errors obtained by both groups of participants are quite similar. Most significant differences were observed in the case of power drill (Figure \ref{fig:powerdrill}) and the spatula. The power drill has a complex geometry and its size also hinders its grasp as the same as spatula and toy airplane. 

Analyzing the overall error in the Figure \ref{chart:error}, we conclude that the largest objects, such as the toy airplane, power drill, and bleach cleaner are those reporting most error rate. In addition, we observe how overall error decreases from the first objects to the last ones. This is mainly because, user skills and expertise with the grasping system are improving progressively. Moreover, results refer to a steep learning curve.

\section{Applications}
\label{sec:applications}

Our grasping system can be applied to several existing problems in different areas of interest, such as: robotics \cite{bric2016current}, rehabilitation \cite{levin2015emergence} and interaction using augmented reality \cite{lv2015touch}.

In robotics, different works have been explored to implement robust grasp approaches that allow robots to interact with the environment. These contributions are organized in mainly four different blocks \cite{bohg2014data}: methods that rely on known objects and previously estimated grasp points  \cite{lin2015robot}, grasping methods for familiar objects\cite{vahrenkamp2016part}, methods for unknown objects based on the analysis of object geometry \cite{zapata2017using} and automatic learning approaches\cite{levine2018learning}. Our approach is more closely related to this last block, where its use would potentially be a relevant contribution. As a direct application, our system enables human-robot knowledge transfer where robots try to imitate human behaviour in performing grasping.

Our grasping system is also useful for rehabilitation of patients with hand motor difficulties, and it could even be done remotely assisted by an expert \cite {escobar2018virtual}, or through an automatic system \cite{avola2018vrheab}. Several works have demonstrated the viability of patient rehabilitation in virtual environments \cite{levin2015emergence}, helping them to improve the mobility of their hands in daily tasks \cite{faria2016benefits}. Our novel error metric in combination with other automatic learning methods, can be used to guide patients during rehabilitation with feedback information and instructions. This will make rehabilitation a more attractive process, by quantifying the patient progress and visualizing its improvements over the duration of rehabilitation.

\begin{figure*}
	\centering
	\begin{subfigure}[b]{0.24\textwidth}
		\includegraphics[width=\textwidth]{Figures/real_0.jpg}
		\caption{RGB image.}
		\label{fig:rgb}
	\end{subfigure}
	\begin{subfigure}[b]{0.24\textwidth}
		\includegraphics[width=\textwidth]{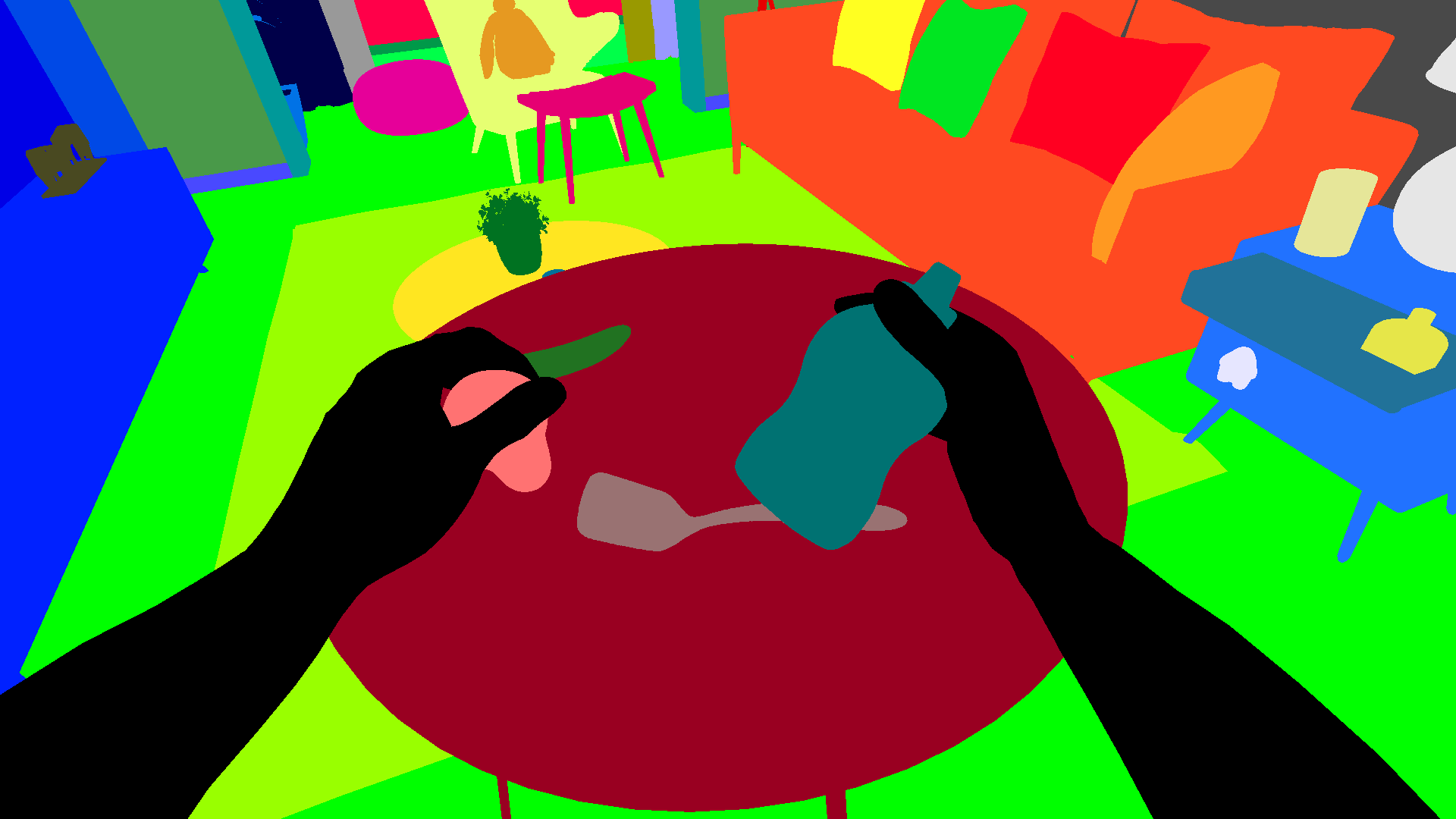}
		\caption{Object mask.}
		\label{fig:mask}
	\end{subfigure}
	\begin{subfigure}[b]{0.24\textwidth}
		\includegraphics[width=\textwidth]{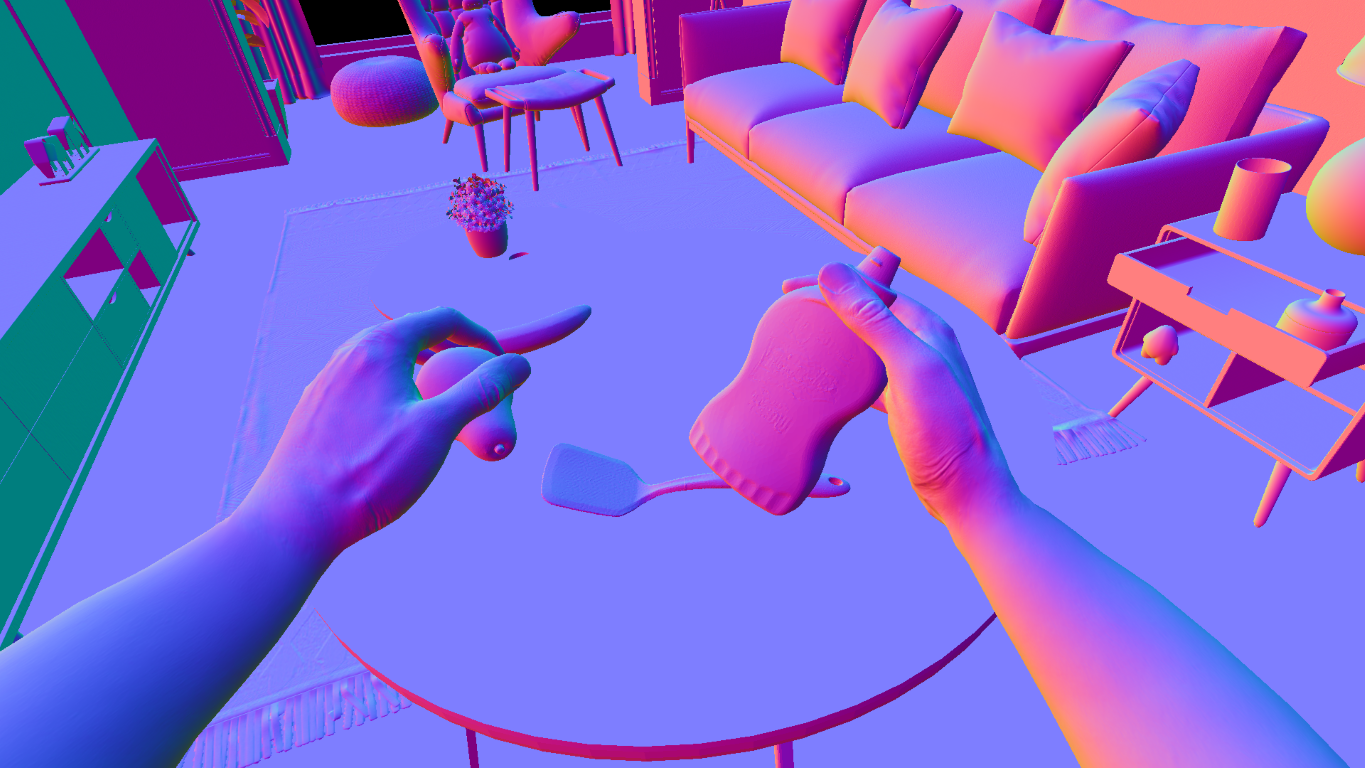}
		\caption{Normal map.}
		\label{fig:normal}
	\end{subfigure}
	\begin{subfigure}[b]{0.24\textwidth}
		\includegraphics[trim={0cm 2cm 0cm 2.1cm},clip,width=\textwidth]{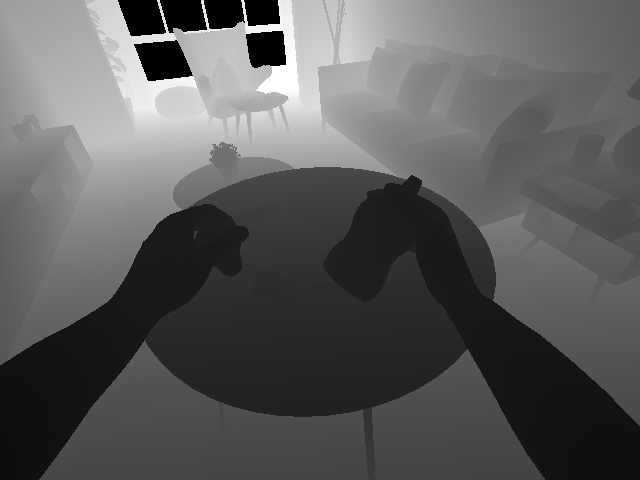}
		\caption{Depth map.}
		\label{fig:depth}
	\end{subfigure}
	\caption{Different ground truth data extracted using UnrealRox alongside our grasping system.}
	\label{fig:unrealroxgt}
\end{figure*}

Finally, our grasping system integrated in UnrealROX \cite{martinez2018} enables many other computer vision and artificial intelligence applications by providing synthetic ground truth data, such as depth and normal maps, object masks, trajectories, stereo pairs, etc. of the virtual human hands interacting with real objects from the YCB dataset (Figure \ref{fig:unrealroxgt}).

\section{Limitations and future works}
\label{sec:limitationsfutureworks}
Our proposal has several major limitations:
\begin{itemize}
	\item Hand movement is based on a single animation regardless object geometry. Depending on the object shape we could vary grasping gesture: spherical-grasping, cylindrical-grasping, finger-pinch, key-pinch, etc. However, our grasping gesture was experimentally the best when dealing with different shaped objects. 
	\item The object can be grasped with only one hand. The user can interact with different objects using both hands at the same time. But not the same object with both hands. 
	\item Sometimes it is difficult to deal with large objects due to the initial hand posture or because objects slide out from the hand palm due to physical collisions. Experienced users can better deal with this problem.
\end{itemize}

As future work, and in order to improve our grasping system, we could vary the hand grip gesture according to the object geometry we are manipulating. This is finding a correspondence between object geometry and a simple shape, e.g. a tennis ball is similar to a sphere thus proceeding with a spherical grasp movement. At the application level, there are several possibilities as we discussed in the previous section. However, we would like to emphasize the use of contact points obtained when grasping an object in virtual reality, to transfer that knowledge and human behavior to real robots.

\section{Conclusion}
\label{sec:conclusion}
This work proposes a flexible and realistic looking grasping system which enables smooth and real-time interaction in virtual reality environments with arbitrary shaped objects. This approach is unconstrained by the object geometry, it is fully controlled by the user and it is modular and easily implemented on different meshes or skeletal configurations. In order to validate our approach, an exhaustive evaluation process was carried out. Our system was evaluated qualitatively and quantitatively by two groups of participants: with previous experience in virtual reality environments (experienced users) and without expertise in VR (inexperienced). For the quantitative evaluation, a new error metric has been proposed to evaluate each grasp, quantifying hand-object overlapping. From the performance analysis results, we conclude that user overall experience was satisfactory and positive. Analyzing the quantitative evaluation, the error difference between experienced users and non experienced is subtle. Moreover, average errors are progressively smaller as more object are grasped. This clearly indicates a steep learning curve. In addition, the qualitative analysis points to a natural and realistic interaction. Users can freely manipulate previously defined dynamic objects in the photorealistic environment. Moreover, grasping contact points can be easily extracted, thus enabling numerous applications, especially in the field of robotics. Unreal Engine 4 project source code is available at GitHub alongside several video demonstrations. This approach can easily be implemented on different game engines.

\appendices

%\section{}

\ifCLASSOPTIONcompsoc
  \section*{Acknowledgments}
\else
  \section*{Acknowledgment}
\fi

This work has been funded by the Spanish Government TIN2016-76515-R grant for the COMBAHO project, supported with Feder funds. This work has also been supported by three Spanish national grants for PhD studies (FPU15/04516, FPU17/00166, and ACIF/2018/197), by the University of Alicante project GRE16-19, and by the Valencian Government project GV/2018/022. Experiments were made possible by a generous hardware donation from NVIDIA.

\ifCLASSOPTIONcaptionsoff
  \newpage
\fi



\bibliographystyle{IEEEtran}
%\bibliography{references}



\begin{IEEEbiography}[{\includegraphics[width=1in,height=1.25in,clip,keepaspectratio]{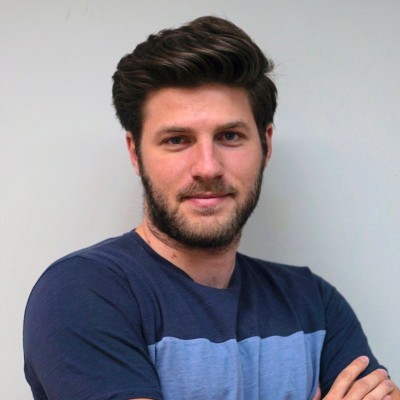}}]{Sergiu Oprea}
is a PhD student in Computer Science at the University of Alicante. He received his Bachelor's Degree in Computer Engineering and his Master's Degree in Automation and Robotics from the same institution in June 2015 and June 2017, respectively. Currently, he is broadly interested in building the next generation of deep learning-based video prediction systems. His main research interests span topics mainly in computer vision, virtual/augmented Reality, and deep learning. He is also a member of European Networks like HiPEAC.	
\end{IEEEbiography}

\begin{IEEEbiography}[{\includegraphics[width=1in,height=1.25in,clip,keepaspectratio]{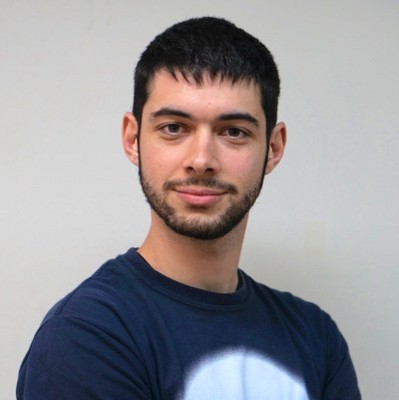}}]{Pablo Martinez-Gonzalez} is a PhD student in Computer Science at the University of Alicante specialized in online deep learning and object pose estimation. He received his Bachelor's Degree in Computer Engineering from the University of Alicante (Spain) in 2015 and his Master's Degree in Computer Graphics, Games and Virtual Reality from the University Rey Juan Carlos (Spain) in 2016. He is also interested in virtual reality and he is a member of European Networks such as HiPEAC.
\end{IEEEbiography}

\begin{IEEEbiography}[{\includegraphics[width=1in,height=1.25in,clip,keepaspectratio]{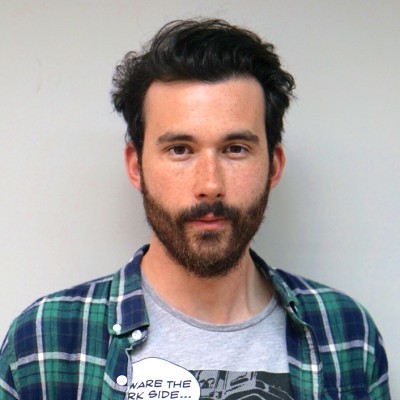}}]{Alberto Garcia-Garcia} is a PhD student (Machine Learning and Computer Vision) at the University of Alicante. He received his Master’s Degree (Automation and Robotics) and his Bachelor’s Degree (Computer Engineering) from the same institution in June 2016 and June 2015 respectively. His main research interests include deep learning (specially graph neural networks), 3D computer vision, and parallel computing on GPUs. He was an intern at Julich Supercomputing Center, intern at NVIDIA working jointly with the Camera/Solutions engineering team and the Mobile Visual Computing group from NVIDIA Research, and intern at Oculus Research (Facebook Reality Labs). He is also a member of European Networks such as HiPEAC and IV\&L.	
\end{IEEEbiography}

\begin{IEEEbiography}[{\includegraphics[width=1in,height=1.25in,clip,keepaspectratio]{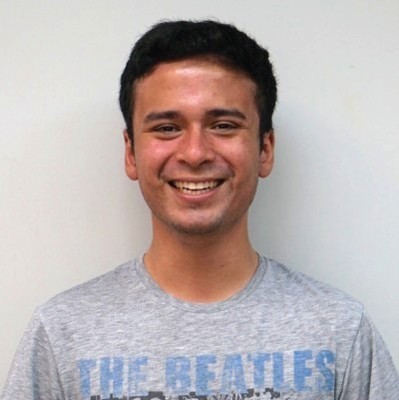}}]{John Alejandro Castro-Vargas} is a PhD student	in Computer Science at the University of Alicante. He received his Bachelor's Degree in Computer Engineering and his Master's Degree in Automation and Robotics from the same institution in June 2016 and June 2017, respectively. His main research interests include deep learning applied to action recognition, robotic grasping and virtual reality. He is also a member of European Networks such as HiPEAC.	
\end{IEEEbiography}

\begin{IEEEbiography}[{\includegraphics[width=1in,height=1.25in,clip,keepaspectratio]{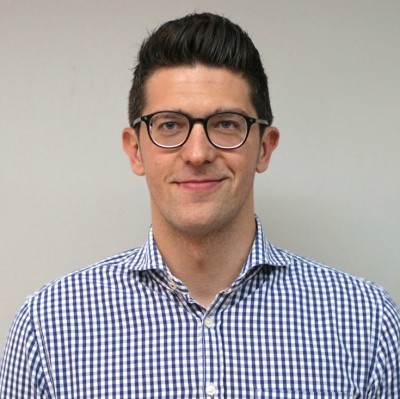}}]{Sergio Orts-Escolano} received a BSc, MSc and PhD in Computer Science from the University of Alicante (Spain) in 2008, 2010 and 2014 respectively. He is currently an assistant professor in the Department of Computer Science and Artificial Intelligence at the University of Alicante. Previously he was a researcher at Microsoft Research where he was one of the leading members of the Holoportation project (virtual 3D teleportation in real-time). His research interests include computer vision, 3D sensing, real-time computing, GPU computing, and deep learning. He has authored +50 publications in journals and top conferences like CVPR, SIGGRAPH, 3DV, BMVC, Neurocomputing, Neural Networks, Applied Soft Computing, etcetera. He is also member of European Networks like HiPEAC and Eucog.
	
\end{IEEEbiography}

\begin{IEEEbiography}[{\includegraphics[width=1in,height=1.25in,clip,keepaspectratio]{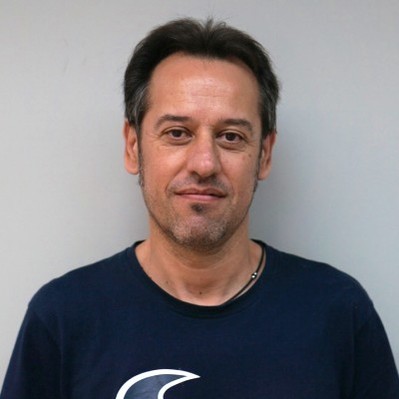}}]{Jose Garcia-Rodriguez} received his Ph.D. degree, with specialization in Computer Vision and Neural Networks, from the University of Alicante (Spain). He is currently Associate Professor at the Department of Computer Technology of the University of Alicante. His research areas of interest include: computer vision, computational intelligence, machine learning, pattern recognition, robotics, man-machine interfaces, ambient intelligence, computational chemistry, and parallel and multicore architectures. He has authored +100 publications in journals and top conferences and revised papers for several journals like Journal of Machine Learning Research, Computational intelligence, Neurocomputing, Neural Networks, Applied Softcomputing, Image Vision and Computing, Journal of Computer Mathematics, IET on Image Processing, SPIE Optical Engineering and many others, chairing sessions in the last decade for WCCI/IJCNN and participating in program committees of several conferences including IJCNN, ICRA, ICANN, IWANN, IWINAC KES, ICDP and many others. He is also member of European Networks of Excellence and COST actions like Eucog, HIPEAC, AAPELE or I\&VL and director or the GPU Research Center at University of Alicante and Phd program in Computer Science.
\end{IEEEbiography}

%\newpage



%\vfill


\end{document}